\documentclass[useAMS,usenatbib,fleqn]{mnras}

\usepackage{amsmath}
\usepackage{bm}
\usepackage{upgreek}

\usepackage{graphicx,color}
\usepackage[normalem]{ulem}

   \usepackage[british]{babel}             
   \usepackage{newtxtext}                  
   \usepackage[slantedGreek]{newtxmath}    
   \usepackage[T1]{fontenc}                

\newcommand{\Exp}[1]{{\rm e}^{#1}}

\newcommand{\eps}{\epsilon}

\newcommand{\bfv}{\bm{v}}

\newcommand{\bfz}{\bm{z}}

\newcommand{\bfs}{\bm{s}}

\newcommand{\f}{_\mathrm{0}}					   	
\newcommand{\1}{_\mathrm{1}}					   	
\newcommand{\CM}{_\mathrm{CM}}					   	
\newcommand{\env}{_\mathrm{env}}					   	
\newcommand{\2}{_\mathrm{2}}					   	

\newcommand{\Ma}{\mathcal{M}}

\newcommand{\soft}{_\mathrm{soft}}

\newcommand{\Gn}{\mathrm{G}}
\newcommand{\init}{_\mathrm{i}}

\newcommand{\acc}{_\mathrm{a}}

\newcommand{\Rsun}{\,\mathrm{R_\odot}}
\newcommand{\Msun}{\,\mathrm{M_\odot}}

\newcommand{\amb}{_\mathrm{amb}}

\newcommand{\CE}{_\mathrm{CE}}

\newcommand{\refine}{_\mathrm{refine}}

\newcommand{\enc}{_\mathrm{enc}}
\newcommand{\boxx}{_\mathrm{box}}
\newcommand{\hyd}{_\mathrm{h}}
\newcommand{\bulge}{_\mathrm{b}}

\newcommand{\cm}{\,{\rm cm}}

\newcommand{\dyn}{\,{\rm dyn}}

\newcommand{\gcmcmcm}{\,{\rm g\,cm^{-3}}}

\newcommand{\K}{\,{\rm K}}

\newcommand{\yr}{\,{\rm yr}}     
\newcommand{\da}{\,{\rm d}}     
\newcommand{\dynecmcm}{\,{\rm dyn\,cm^{-2}}}

\def\apj{ApJ}

\def\mnras{MNRAS}
\def\aap{A\&A}
\def\apjl{ApJ}

\def\apjs{ApJS}

\title[Drag Force in Global Common Envelope Simulations] 
{How Drag Force Evolves in Global Common Envelope Simulations}
\author[L.~Chamandy et al.]{Luke Chamandy,\thanks{lchamandy@pas.rochester.edu}
Eric G.~Blackman, 
Adam Frank, 
Jonathan Carroll-Nellenback, 
\newauthor 
Yangyuxin Zou
and Yisheng Tu 
\\
Department of Physics and Astronomy, University of Rochester, Rochester NY 14627, USA\\
}

\begin{document}


\maketitle

\begin{abstract} 
We compute the forces, torque and rate of work on the companion-core binary due to drag in global simulations of common envelope (CE) evolution for three different companion masses.  Our simulations help to delineate regimes when conventional analytic drag force approximations are applicable.  During and just prior to the first periastron passage of the in-spiral phase, the drag force is reasonably approximated by conventional analytic theory and peaks at values proportional to the companion mass.  Good agreement between global and local 3D ``wind tunnel'' simulations, including similar net drag force and flow pattern, is obtained for comparable regions of parameter space.  However, subsequent to the first periastron passage, the drag force is up to an order of magnitude smaller than theoretical predictions, quasi-steady, and depends only weakly on companion mass.  The discrepancy is exacerbated for larger companion mass and when the inter-particle separation reduces to the Bondi-Hoyle-Lyttleton accretion radius, creating a turbulent thermalized region.  Greater flow symmetry  during this phase leads to near balance of opposing gravitational forces in front of and behind the companion, hence a small net drag.  The reduced drag force at late times helps explain why companion-core separations necessary for envelope ejection are not reached by the end of limited duration CE simulations.
\end{abstract}

\defcitealias{Chamandy+18}{Paper~I}
\defcitealias{Chamandy+19a}{Paper~II}
\defcitealias{Macleod+17}{M17}
\defcitealias{Dodd+Mccrea52}{DM}


\section{Introduction} \label{sec:intro}

Common envelope evolution (CEE) is the most natural mechanism for rapidly tightening binary orbits and  likely facilitates many phenomena, including gravitational wave-emitting mergers and type Ia supernovae.
In CEE, the primary and secondary cores inspiral from drag,
transferring orbital energy to the envelope until the latter ejects,  or the cores merge.

Hydrodynamic simulations of this process generally do not eject the envelope.
Although it is possible that the cores should merge for the parameter regime 
explored in some of the simulations \citep{Iaconi+18},
in other simulations the rate of decay of the inter-particle separation $a$ decreases dramatically 
at values of $a$ too large for a merger during the computation.
There may also be missing physics in the simulations.
For example, most  employ an ideal gas equation of state (EOS), 
whereas a more sophisticated EOS should account for ionization and recombination. 
When recombination energy is injected locally, 
the envelope is found to eject or almost eject in at least some cases \citep{Nandez+15,Nandez+Ivanova16,Prust+Chang19}.

\citet{Chamandy+19a} (hereafter \citetalias{Chamandy+19a}) applied the CE energy formalism \citep{Vandenheuvel76,Webbink84,Livio+Soker88} 
to show that for a reasonable energy parameter $\alpha\CE\lesssim0.3$, 
theory correctly predicts that the envelope will not eject in our simulation 
or in any other with very similar initial conditions \citep{Ohlmann+16a} 
because the simulations do not reach the predicted separation for ejection by the end of the runs.
That simulations do not eject the envelope because they do not attain small enough separations is partly supported  by observations,
exhibiting small  final separations \citep{Iaconi+17,Iaconi+Demarco19}.

Although extra energy sources (e.g. recombination energy or energy released by accretion onto the companion) 
may help to eject the envelope,
they should also  result in  larger final separations,
since less transfer of orbital energy  is then  required for ejection.
On the other hand, energy sinks, such as loss via radiation,  may offset energy gain by the envelope gas
(\citealt{Sabach+17}; \citealt{Grichener+18}, but see \citealt{Ivanova18}).

Separations at late times tend to be overestimated because of inadequate numerical resolution 
(\citealt{Ohlmann+16a,Iaconi+17,Iaconi+18}; \citetalias{Chamandy+19a}),
but this is unlikely a dominant effect--the slow decrease of $a$ at late times needs to be explained physically.
\citet{Ricker+Taam08,Ricker+Taam12,Staff+16b} and \citet{Iaconi+17,Iaconi+18}
include some explorations of the drag force in their global CE simulations.
And recent work by \citet{Reichardt+19} showed that the decrease in the rate of orbital tightening at late times 
was consistent with reduction in the drag force on the companion measured in one of their simulations. 
The reduction was explained qualitatively by a reduction 
in the angular velocity of the cores relative to the gas in their vicinity.

The goal of this work is to analyze the drag force in three otherwise identical simulations, 
but each with a different companion mass, 
and to compare our results with results from analytic theory 
and local wind tunnel CE simulations of flow near the secondary.
In Sec.~\ref{sec:methods} we summarize our numerical methods.
Sec.~\ref{sec:results} contains the results of our simulations for the net force. 
We compare these results to analytic theory in Sec.~\ref{sec:comparison_analytic}.
The evolution of the flow around the secondary, 
with a focus on the simulation with largest companion mass, is explored in Sec.~\ref{sec:2D}.
The results for the net force and flow properties are then compared to wind tunnel simulations 
in Sec.~\ref{sec:comparison_windtunnel}.
We summarize and conclude in Sec.~\ref{sec:conclusions}.

\section{Simulation parameters and methods} \label{sec:methods}
We employ the hydrodynamics code \textsc{astrobear}, which includes adaptive mesh refinement (AMR).
The primary is an $M\1=1.96\Msun$ red giant branch (RGB) star with radius $R\1=48\Msun$ and core mass $M_\mathrm{1,c}=0.37\Msun$,
and the secondary has mass $M\2=0.98\Msun$ (Model~A), $M\2=0.49\Msun$ (Model~B) or $M\2=0.245\Msun$ (Model~C).
The primary and secondary are initialized in a circular orbit with separation $a\init=49\Rsun$.
Aside from the companion mass and initial velocities, the three runs are identical.
Model~A is the same as Model~A of \citet{Chamandy+18} (hereafter \citetalias{Chamandy+18}) and \citetalias{Chamandy+19a}.

RGB core and companion are modeled as point particles (``particle~1'' and ``particle~2'', respectively)
that interact with each other and gas via gravity only.
The particle potential is smoothed according to a spline function \citep{Springel10}
such that it is Newtonian for $r>r\soft$ and shallower than Newtonian for $r<r\soft$, where $r\soft$ is the spline softening radius.
The RGB model is adapted from a \textsc{mesa} \citep{Paxton+15} 1D profile using a similar method to that of \citet{Ohlmann+17} 
to model the gas profile within the softening radius.
The spline softening radius and  smallest resolution element are respectively 
$r\soft=2.4\Rsun$ and $\delta=0.14\Rsun$ from $t=0$ to $t=16.7\da$, and $r\soft=1.2\Rsun$ and $\delta=0.07\Rsun$ thereafter.
Refinement at the highest resolution is applied \textit{everywhere} within a sphere of dynamically changing radius $r\refine$
(see Fig.~\ref{fig:separation}), centred on the primary core before $t=16.7\da$, and companion thereafter.
The simulation domain size is $L\boxx=1150\Rsun$, with $512^3$ base cells of size $2.25\Rsun$.
(Four levels of AMR are used for $t<16.7\da$ and five levels thereafter, and going up one level halves the cell size.).
Extrapolation boundary conditions are employed.

An ideal gas EOS with $\gamma=5/3$ is employed.
The ambient density and pressure are $\rho\amb=6.7\times10^{-9}\gcmcmcm$ and $P\amb=1.0\times10^5\dynecmcm$.
The simulations are stopped after $t=40\da$.
More details about the setup and methods can be found in Papers~I and II.

In Model~B of \citetalias{Chamandy+18}, the secondary was a sink particle that accreted mass 
at a rate which was an upper bound to the true accretion rate. 
Since the orbit, and hence the drag force, were not drastically affected by this accretion,
we exclude accretion onto the companion in the present simulations.

\section{Overall Evolution}\label{sec:results}

\subsection{Orbital separation}
\begin{figure}
  \vspace{1cm} 
  \includegraphics[width=\columnwidth,clip=true,trim= 8 0 7 0]{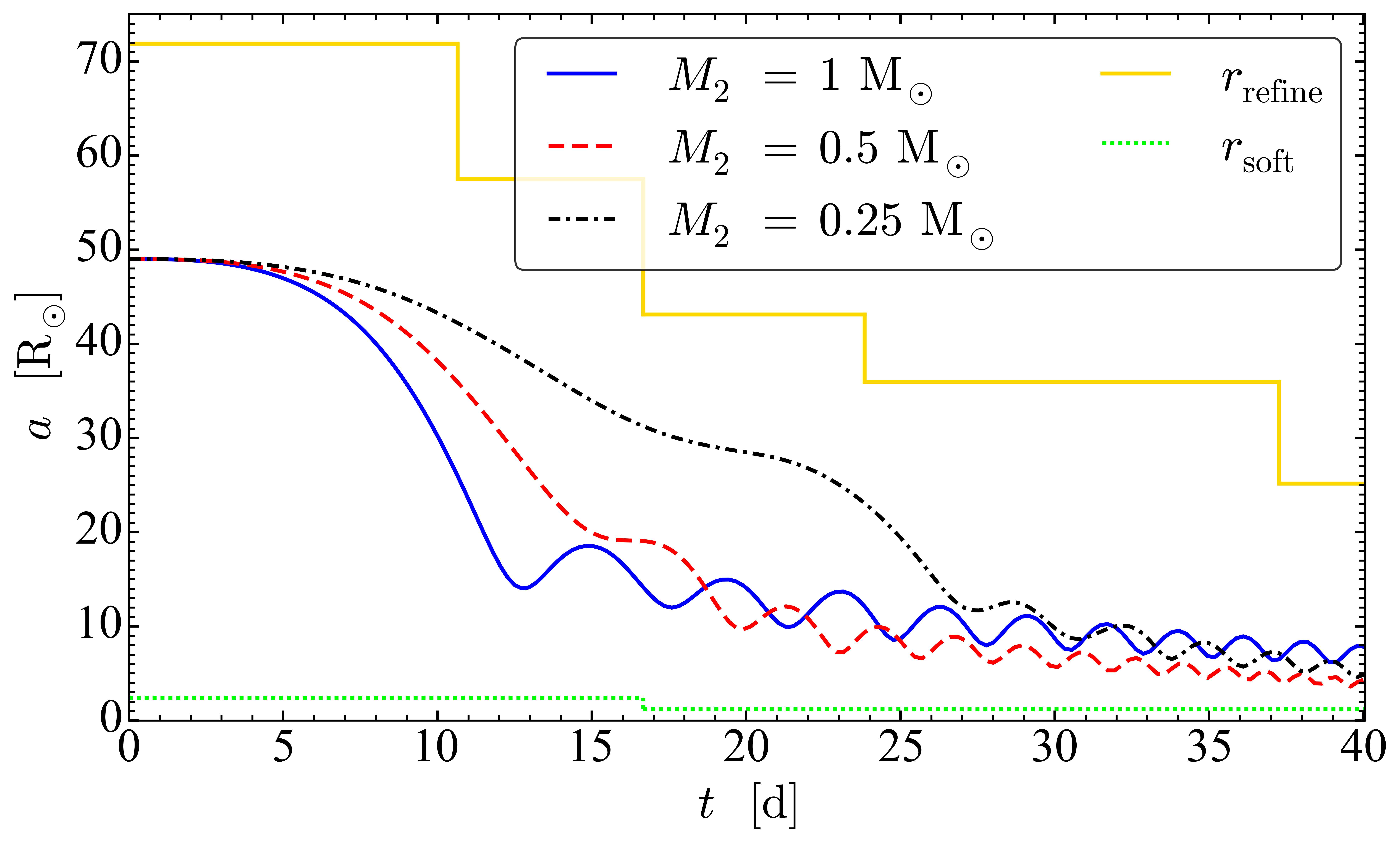}
  \caption{Inter-particle separation as a function of time for the three runs.
           \label{fig:separation}
          }
\end{figure}
 
Fig.~\ref{fig:separation} shows orbital separation  versus time 
for Models~A, B and C in solid blue, dashed red and dash-dotted black, respectively.
The quantities $r\soft$ and $r\refine$, which do not change between runs, are also shown for reference.
As the companion mass is lowered, the initial orbital speed and separation decay rate are both reduced.
At later times however, the separation decays more rapidly for lower mass, and the curves  cross.
This behaviour is consistent with other studies \citep[e.g.][]{Passy+12b}.

We have computed the tidal shredding radius $r_\mathrm{shred}$ for a main sequence secondary using the initial density profile of the primary
along with the estimate of \citet{Nordhaus+Blackman06} and the mass-radius relation from \citet{Eker+18}, 
and find $r_\mathrm{shred}<1\Rsun$ for all three models.
For a white dwarf secondary, $r_\mathrm{shred}$ would be smaller still.
Thus, the secondary is not expected to tidally shred during any  of  our  simulation runs.

\begin{figure*}
\begin{center}
  \includegraphics[width=0.9\textwidth,clip=true,trim= 0 40 0 0]{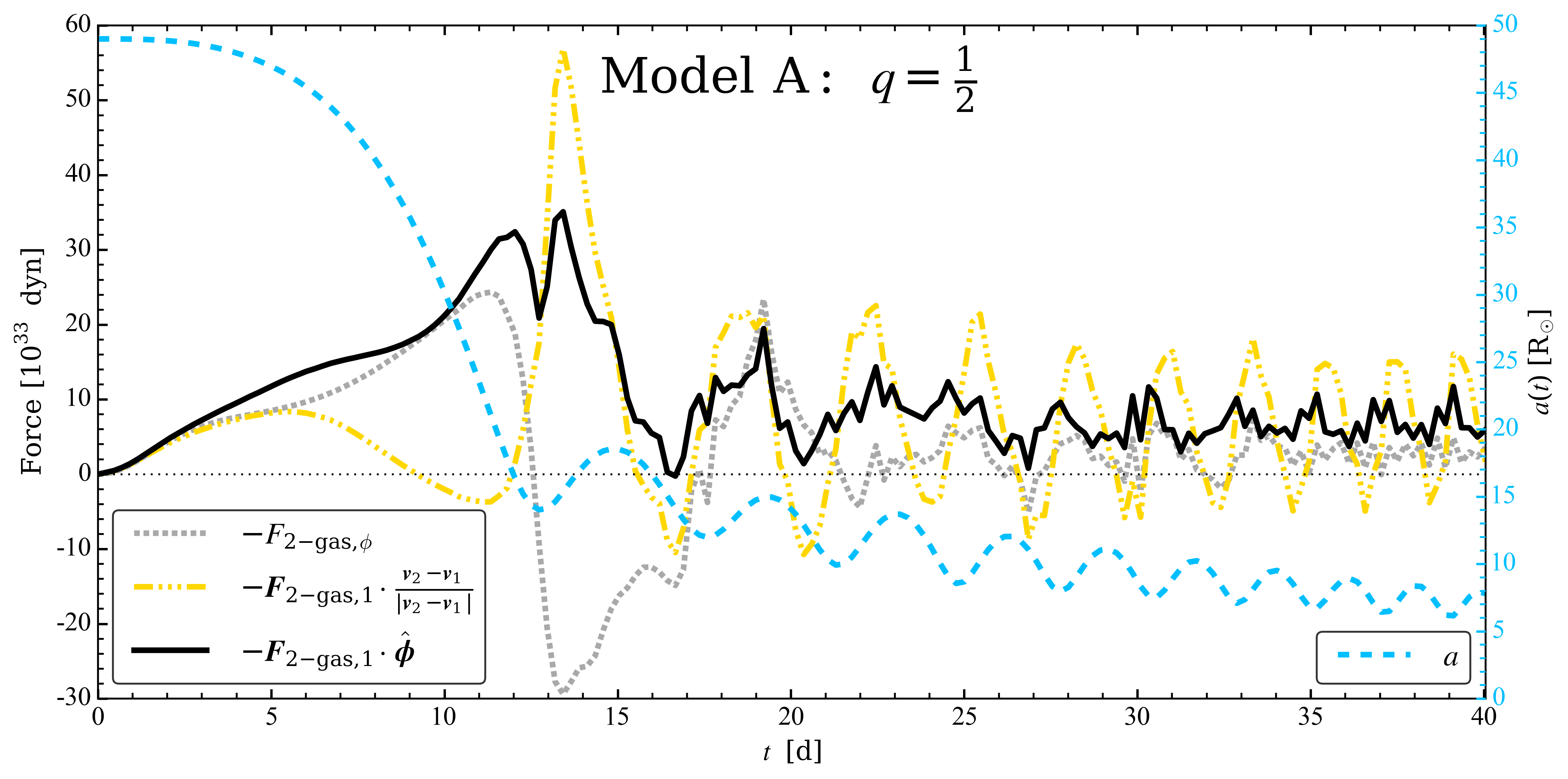}\\
  \includegraphics[width=0.9\textwidth,clip=true,trim= 0 40 0 0]{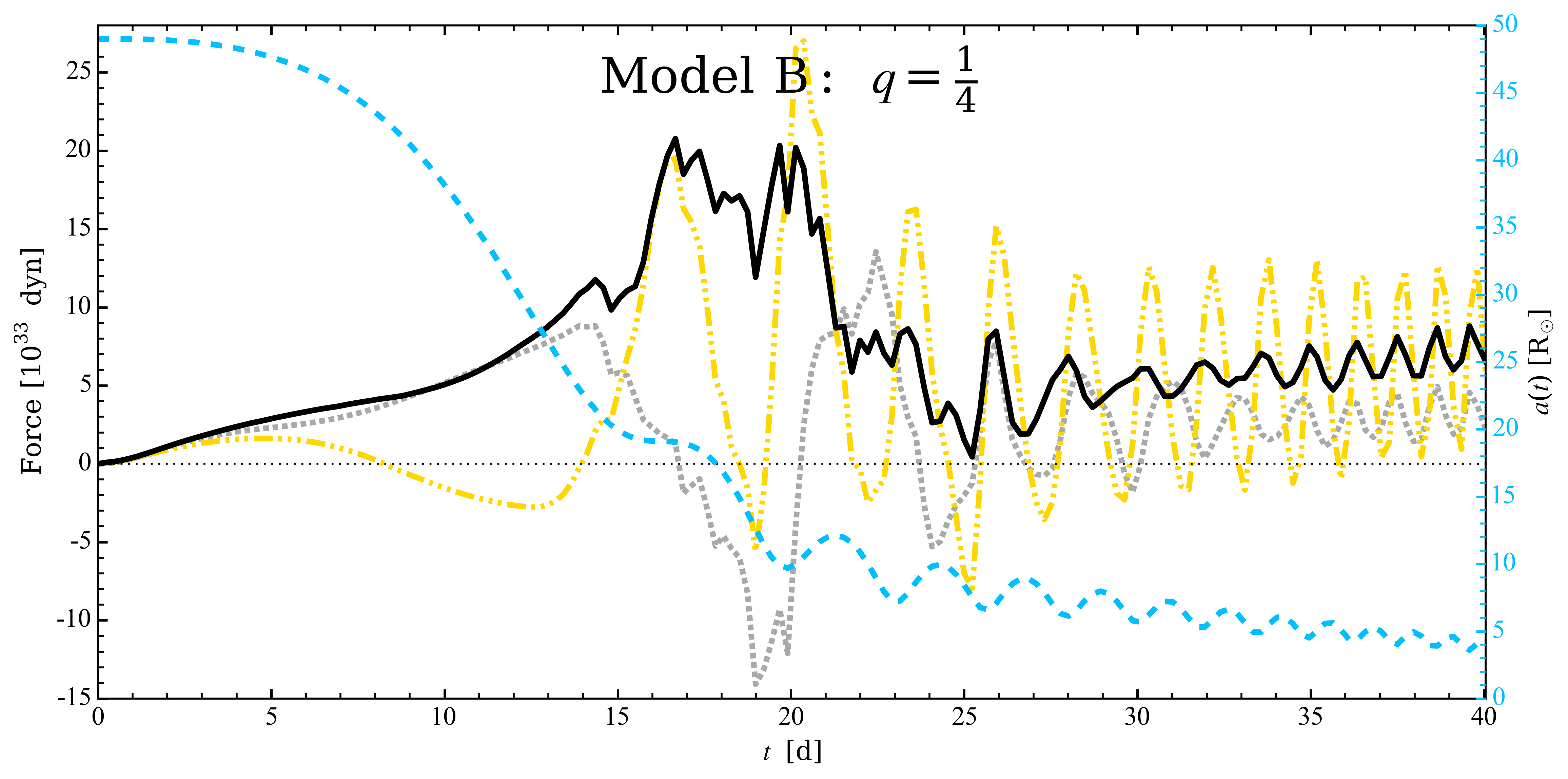}\\
  \includegraphics[width=0.89\textwidth,clip=true,trim= 0 0 4 0]{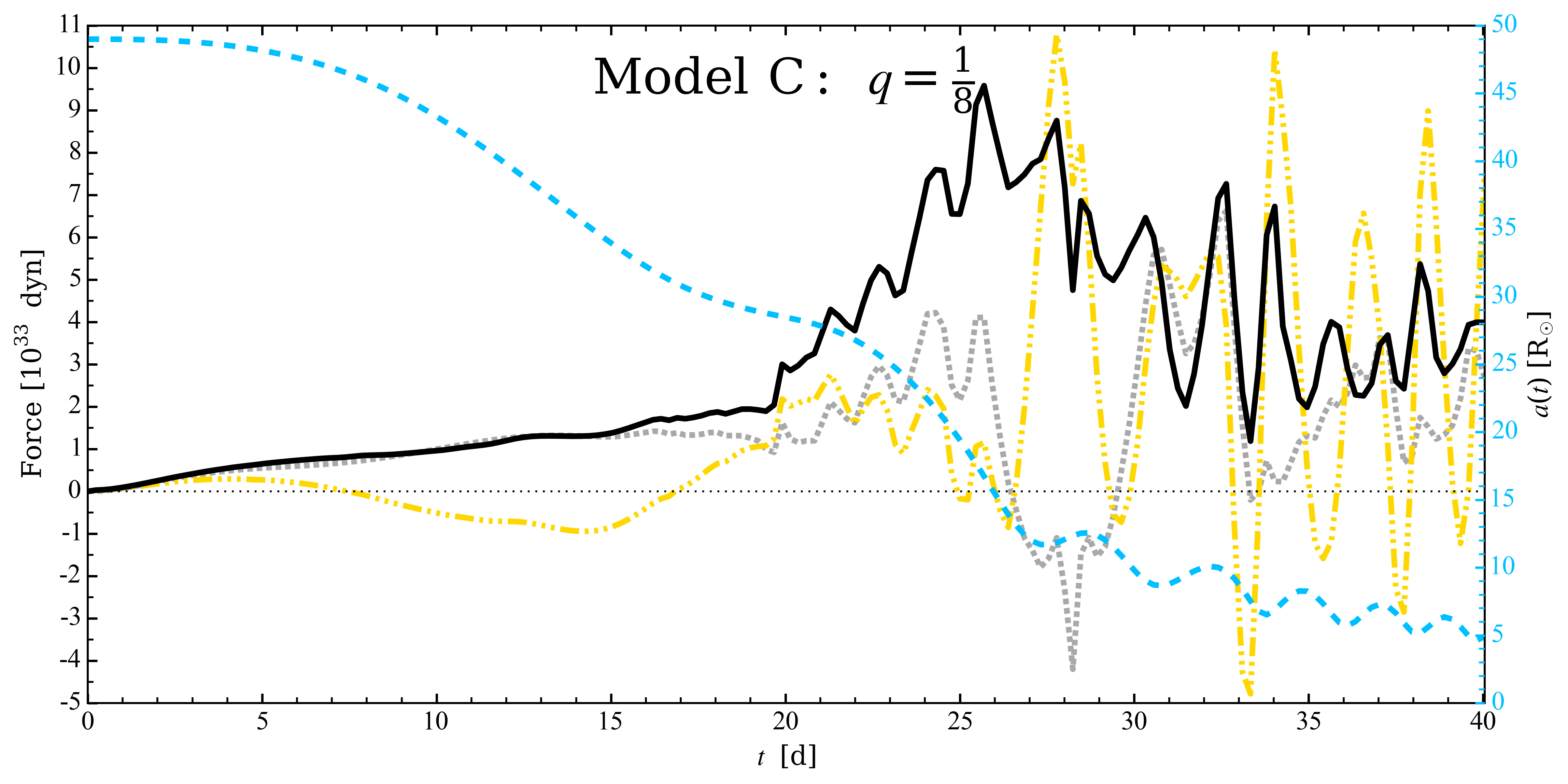}
  \caption{Azimuthal ($\phi$) component of the net force on particle~2 
           due to the gas in the non-inertial rest frame of particle~1, 
           computed from the simulation (solid black), 
           component of this force along the relative velocity of particle~2 with respect to particle~1 (dash-triple-dotted gold),
           and contribution to the $\phi$-component from the force on particle~2 in the lab frame, without the fictitious force (dotted grey).
           The inter-particle separation (dashed light blue) is plotted using the right axis, for reference.
           \label{fig:force2}
          }            
\end{center}
\end{figure*}

\subsection{Drag Force}

The centre of mass of the particles accelerates during the simulation \citepalias{Chamandy+19a}
due to the gravitational interaction between the gas and each of the particles.
To facilitate comparison with theory and local simulations that treat the primary as fixed and non-rotating, 
we compute the dynamical friction force on particle~2 in the non-inertial rest frame 
of particle~1, $\bm{F}_\mathrm{2-gas,1}$,
where the subscript `1' after the comma denotes this reference frame. 
As seen in the lab frame, this frame orbits with particle~1 but does not rotate.
This introduces a fictitious force so that the force exerted on particle~2 by gas in this frame is given by
\begin{equation}
  \label{F2gas_frame1}
  \bm{F}_\mathrm{2-gas,1}= \bm{F}_\mathrm{2-gas} -(M\2/M_\mathrm{1,c})\bm{F}_\mathrm{1-gas},
\end{equation}
where the terms on the right  are computed in the lab frame (nearly the centre of mass frame of the entire system; see \citetalias{Chamandy+19a}).
We have not included the terms $(1+M\2/M_\mathrm{1,c})\bm{F}_{2-1}$ because these terms involve forces between the particles,
and we are interested in computing the forces between gas and particles.
Note that a force in the $-\phi$ direction (a drag) on particle~1 in the lab frame 
contributes a drag on particle~2 in the frame of particle~1. 
To compute the terms on the right  of equation~\eqref{F2gas_frame1}, 
we simply integrate the force per unit volume on each particle, for example: 
$\bm{F}_\mathrm{2-gas}= \Gn M\2 \sum_V \rho(\bfs)[(\bfs-\bfs\2)/|\bfs-\bfs\2|^3]\mathrm{d}^3s$,
where $\rho(\bfs)$ is the gas density at position $\bfs$, 
$V$ is the volume of the simulation domain, and $\bfs\2$ is the position of particle~2.
We then compute the $\phi$-component $(\bfs\2-\bfs\1)\times\bm{F}_\mathrm{2-gas,1}/a\cdot\hat{\bfz}$.
Likewise, we compute the projection of the force along the velocity vector relative to particle~1: 
$\bm{F}_\mathrm{2-gas,1}\cdot(\bfv\2-\bfv\1)/|\bfv\2-\bfv\1|$.

The force on particle~2, multiplied by $-1$, is presented in Fig.~\ref{fig:force2}.
We refer to positive values on the plot as `drag' and negative values as `thrust'.
The $\phi$-component is plotted as a solid black line,
and the projection along the relative velocity is plotted as a dash-triple-dotted gold line,
for Model~A (top), Model~B (middle) and Model~C (bottom).
The separation $a$ is plotted with respect to the right axis for reference.%
\footnote{The inter-particle separation never differs from its projection in the $xy$-plane by more than $0.2\%$.}

For all models, the $-\phi$-component of the force steadily increases from $t=0$, 
attains a broad peak of a few days width (with a double humped morphology, at least for Models~A and B),
and then reduces  before  becoming  quasi-constant until the end of the simulation.
The broad peak roughly coincides with the first periastron passage, 
though for Models~B and C it happens slightly earlier. The peak magnitude is roughly proportional to the companion mass: $\sim32$, $\sim16$ and $\sim8$ in units of $10^{33}\dyn$ for Models~A, B and C, respectively.

Periodicity emerges at later times,  particularly in Models~A and B, with the force magnitude greatest (smallest) when $a$ is smallest (greatest).
The  evolution is slower in Model~C, so we  expect such variations to become more regular only after $t=40\da$.
At late times, the magnitude is only weakly dependent on the companion mass, being $\sim7\times10^{33}\dyn$ for Models~A and B and closer to $\sim4\times10^{33}\dyn$ for Model~C, but has not yet stabilized by the end of that run.

The dotted grey curve shows the contribution to the solid black curve from 
only the  the first term on the right of equation~\eqref{F2gas_frame1}.
\textit{Ignoring the  fictitious force exerted by gas on particle~1 would thus lead to the wrong conclusion 
that the $\phi$-component of $\bm{F}_\mathrm{2-gas,1}$ is sometimes positive.}

The component of  force along $\bfv\2-\bfv\1$ has a period-averaged magnitude similar to the $\phi$-component at late times, but varies strongly with orbital separation and oscillates between drag and thrust. The oscillations occur because the  gas force exerted on each particle is dominated by gas in the direction of the other particle, so particle approach (recession) produces thrust (drag).

To explore the effect of changing the softening length and resolution at $t=16.7\da$,
we compared the drag force in Model~A to that of Model~F from \citetalias{Chamandy+19a},
for which both $r\soft$ and $\delta$ retain their initial values for the full $t=37.3\da$ simulation.
The separation $a(t)$ for Model~F differs only slightly from that of Model~A, as does the force  (Appendix~\ref{sec:152}).

\begin{figure*}
\begin{center}
  \includegraphics[width=0.495\textwidth,clip=true,trim= 0 30 0 0]{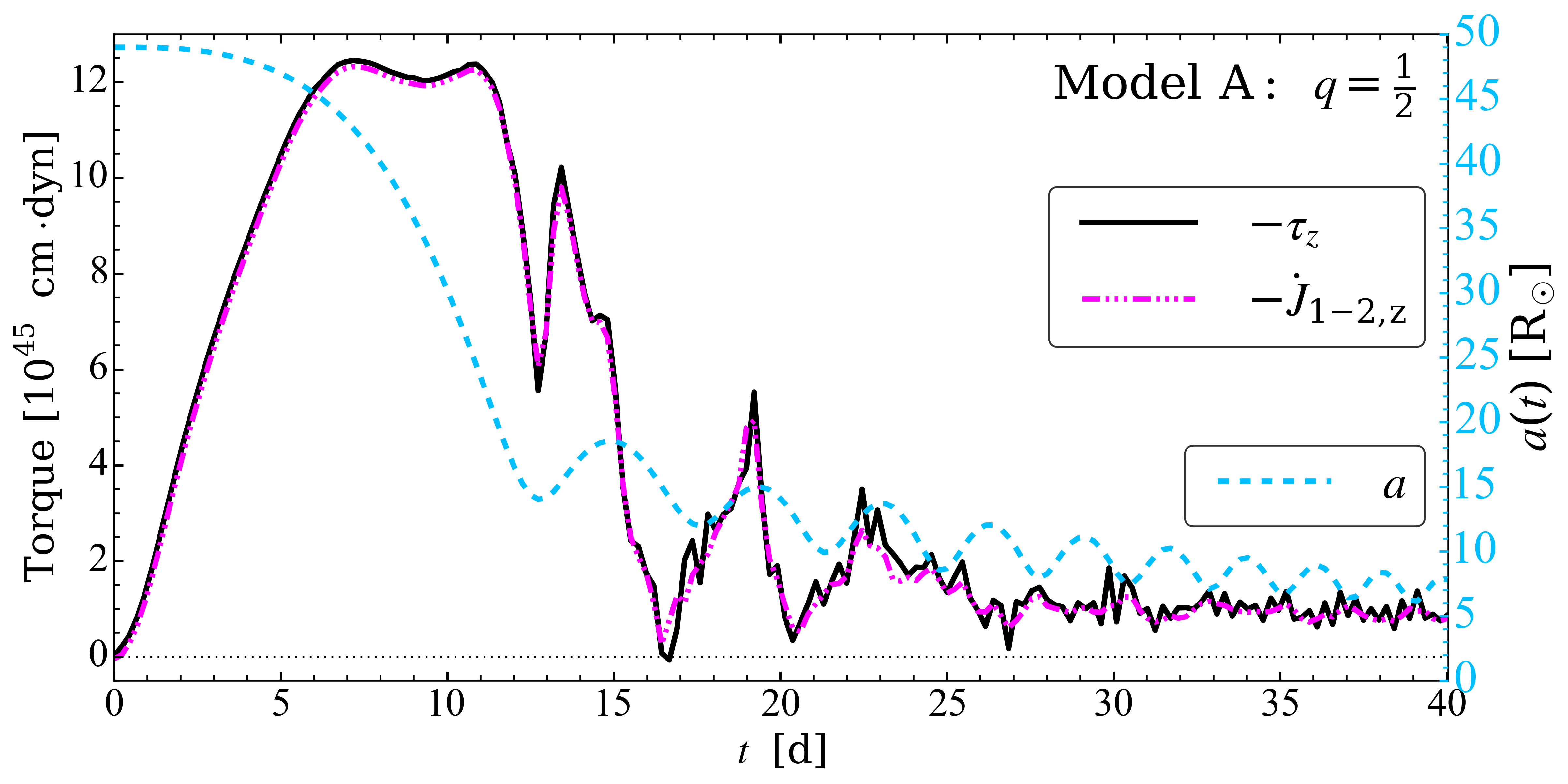}
  \includegraphics[width=0.495\textwidth,clip=true,trim= 0 30 0 0]{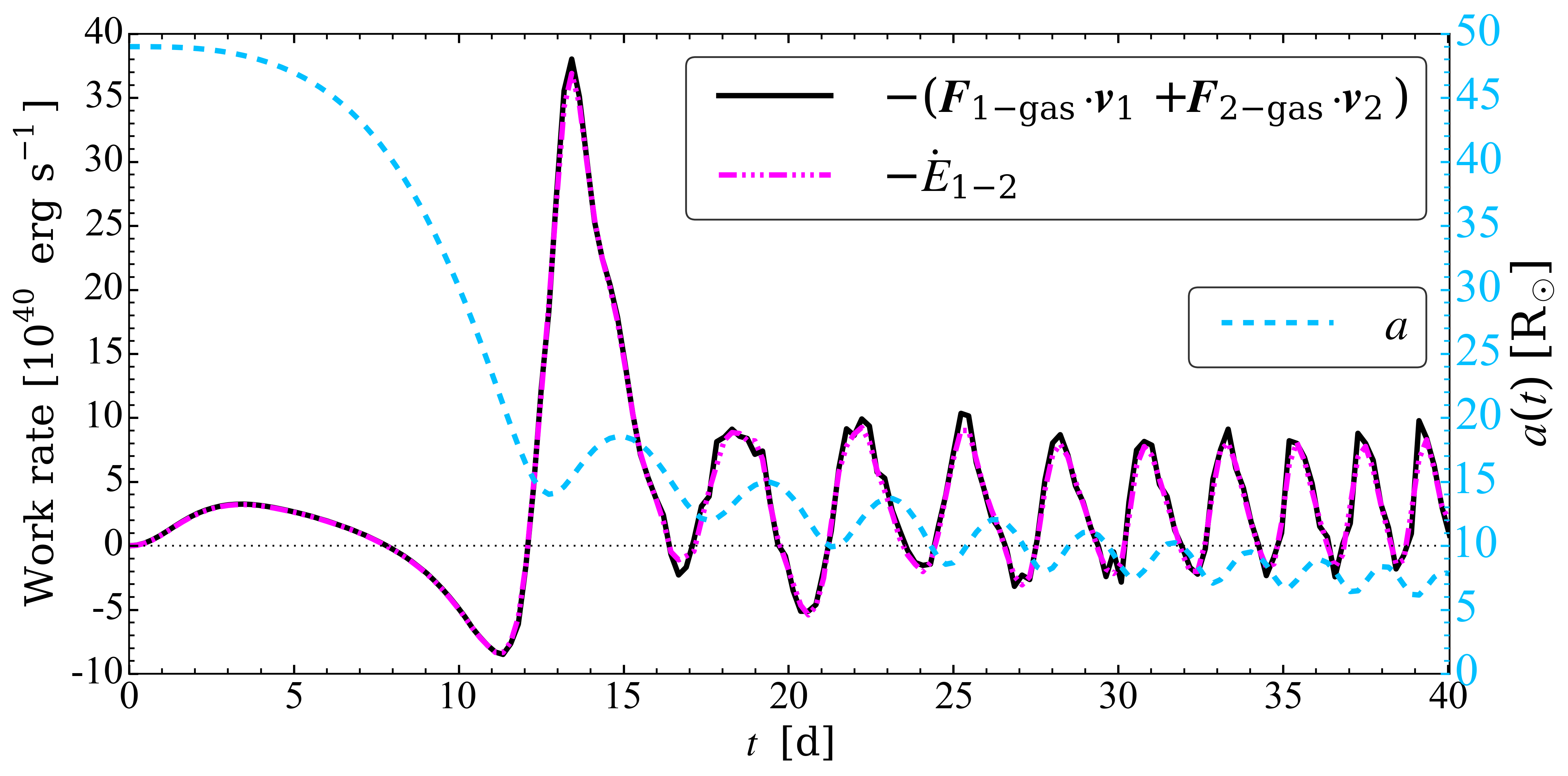}\\
  \includegraphics[width=0.495\textwidth,clip=true,trim= 0 30 0 0]{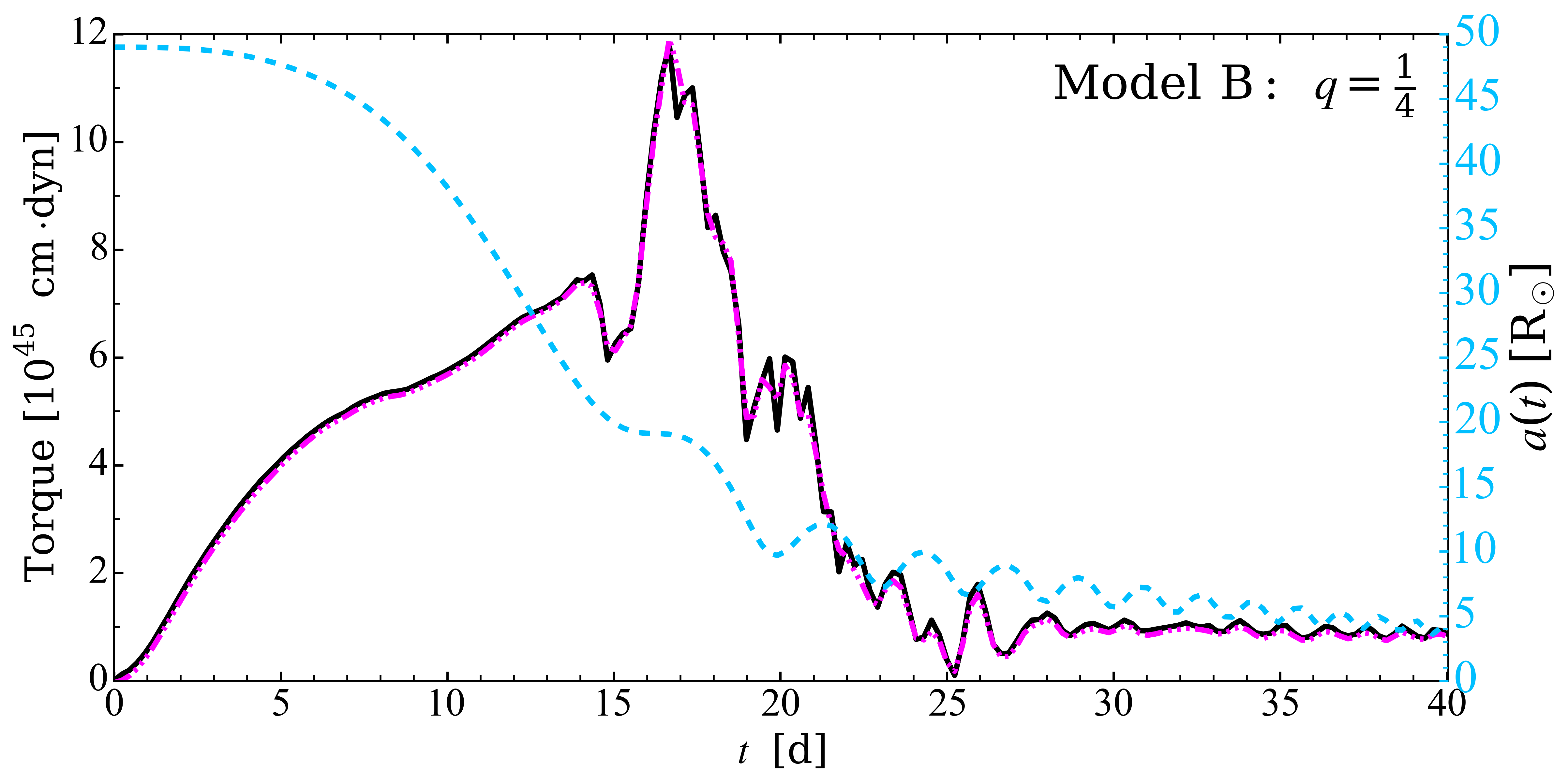}
  \includegraphics[width=0.495\textwidth,clip=true,trim= 0 30 0 0]{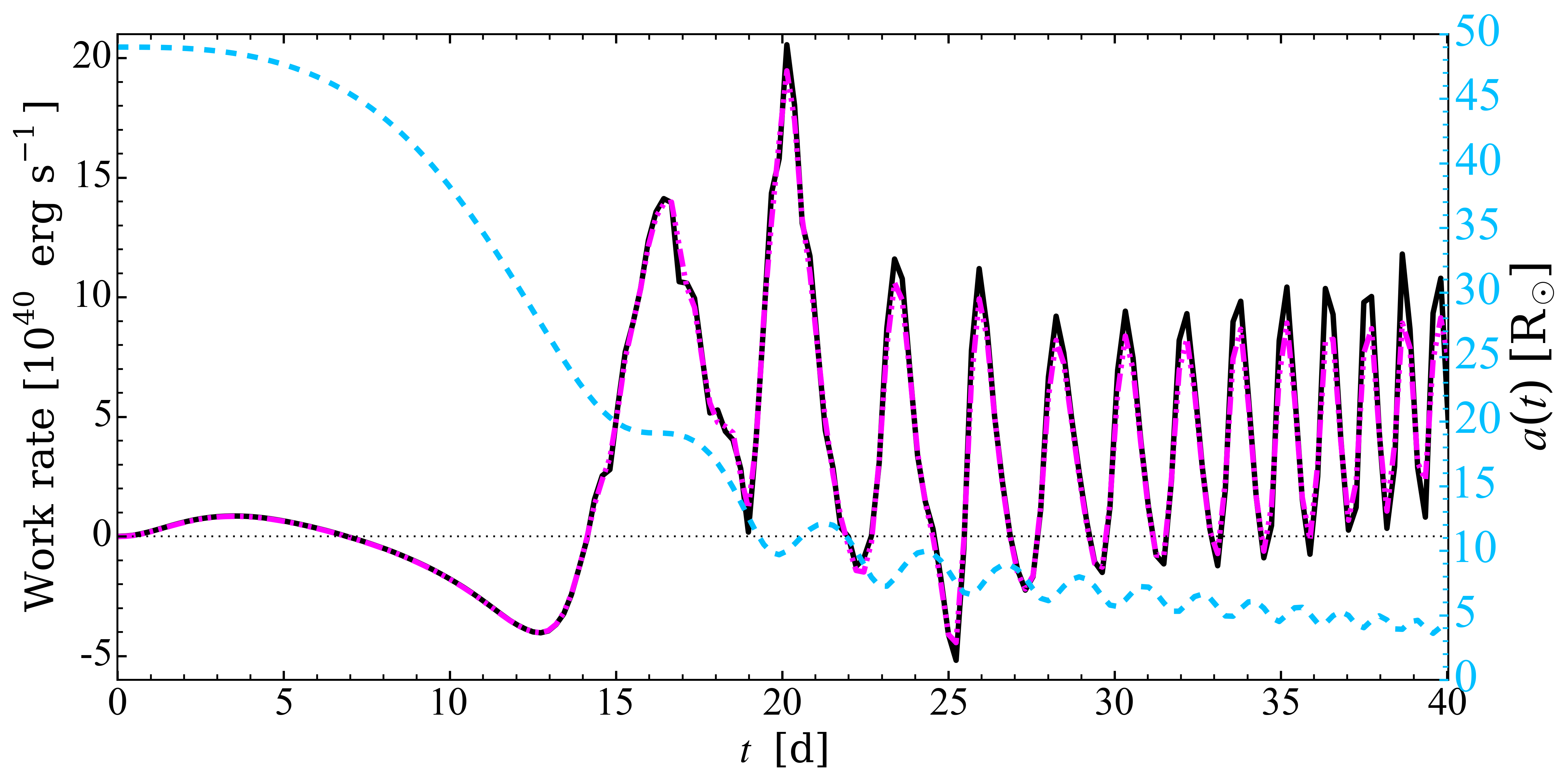}\\
  \includegraphics[width=0.495\textwidth,clip=true,trim= 0  0 0 0]{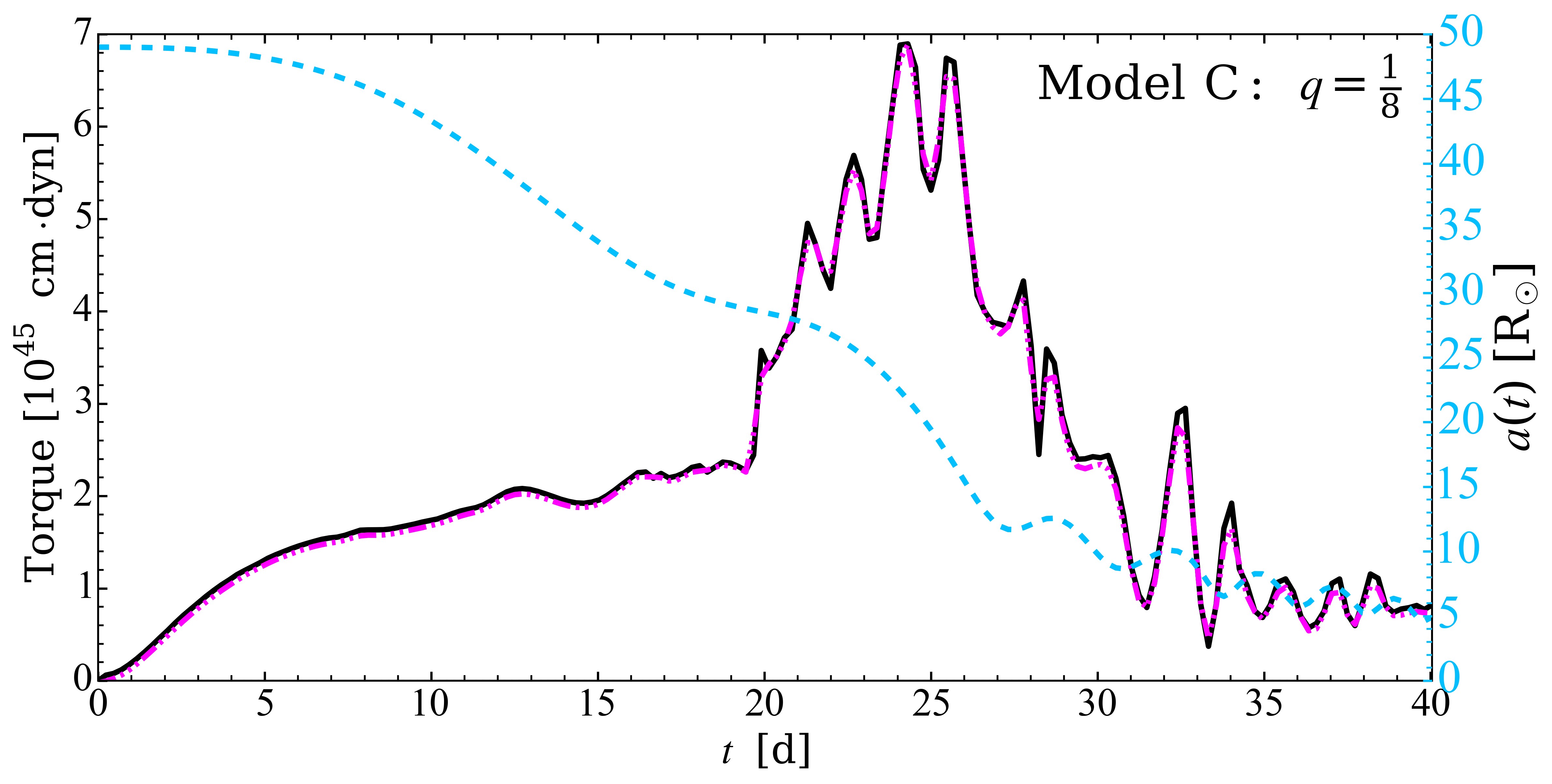}
  \includegraphics[width=0.495\textwidth,clip=true,trim= 0  0 0 0]{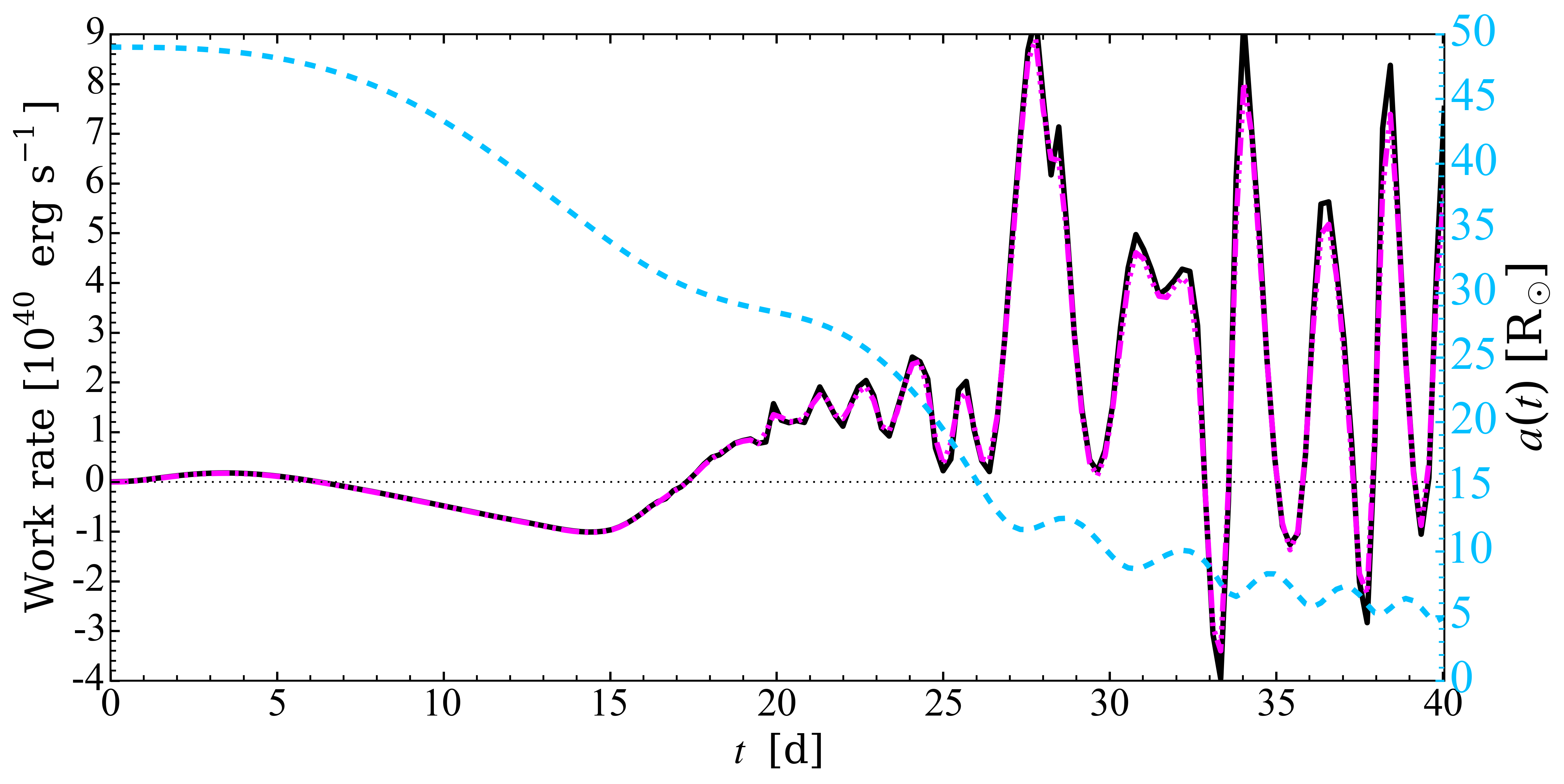}\\
  \caption{Left: Torque on particles about the particle centre of mass.
           The torque computed from the forces is shown solid black, 
           while that computed from the rate of change of the particle angular momentum is shown dash-triple-dotted magenta.
           Right: Similar to left panels but now showing the rate of change of work done by gas on particles
           in the inertial frame, computed from the forces or the rate of change of the orbital energy.
           \label{fig:torque_luminosity}
          }            
\end{center}
\end{figure*}

\subsection{Torque and Orbital Energy Dissipation} \label{sec:torque}
The torque on the particles about their centre of mass is 
plotted on the left side of Fig.~\ref{fig:torque_luminosity}.
The solid black line shows the $z$-component of the torque computed using the forces obtained by integration over the simulation domain,
\begin{equation}
  \tau_z= \frac{M\2}{M_\mathrm{1,c}+M\2}aF_\mathrm{1-gas,\phi} +\frac{M_\mathrm{1,c}}{M_\mathrm{1,c}+M\2}aF_\mathrm{2-gas,\phi}.
\end{equation}
Here the forces  are in the lab frame because torques  from fictitious forces cancel.

The dash-triple-dotted magenta line shows the $z$-component of the rate of change of the particle angular momentum.
This is obtained by first computing
the angular momentum of the particles about the particle centre of mass (denoted `CM')
\begin{equation}
\begin{split}
  J_\mathrm{1-2,z}=&\,M_\mathrm{1,c}[(\bfs\1-\bfs\CM)\times(\bfv\1-\bfv\CM)]_z\\ 
      &+M\2[(\bfs\2-\bfs\CM)\times(\bfv\2-\bfv\CM)]_z,
\end{split}
\end{equation}
and then numerically time-differentiating $J_z$  (with sampling interval $\approx 0.23\da$ per frame).
The two methods of computation should in principle yield identical results,
except for the sampling error on $\dot{J}_z$, and the level of agreement is indeed excellent. 
The $xy$-plane components of the torque are negligible in both methods.

Given that the particle orbital energy dissipation rate could be used 
to estimate the observed luminosity of potential transient CE events such as luminous red novae, we also  compute this. 
However, such an estimate would need to consider radiative transfer and is left for future work.
The right side of Fig.~\ref{fig:torque_luminosity} shows  strong agreement between the two different methods, 
first from computing the rate of work done by gas on particles,
\begin{equation}
  \dot{W}= \bm{F}_\mathrm{1-gas}\cdot\bm{v}\1 +\bm{F}_\mathrm{2-gas}\cdot\bm{v}\2
\end{equation}
(black solid line),
and second by numerical time-differentiating the total particle energy 
\begin{equation}
  E_\mathrm{1-2}= \frac{1}{2}M_\mathrm{1,c}v_1^2 +\frac{1}{2}M\2v_2^2 -\frac{\Gn M_\mathrm{1,c}M\2}{a}
\end{equation}
(dash-triple-dotted magenta line).

\section{Comparison to Analytic Theory} \label{sec:comparison_analytic}

\begin{figure*}
\begin{center}
  \includegraphics[width=0.9\textwidth,clip=true,trim= 0 40 0 0]{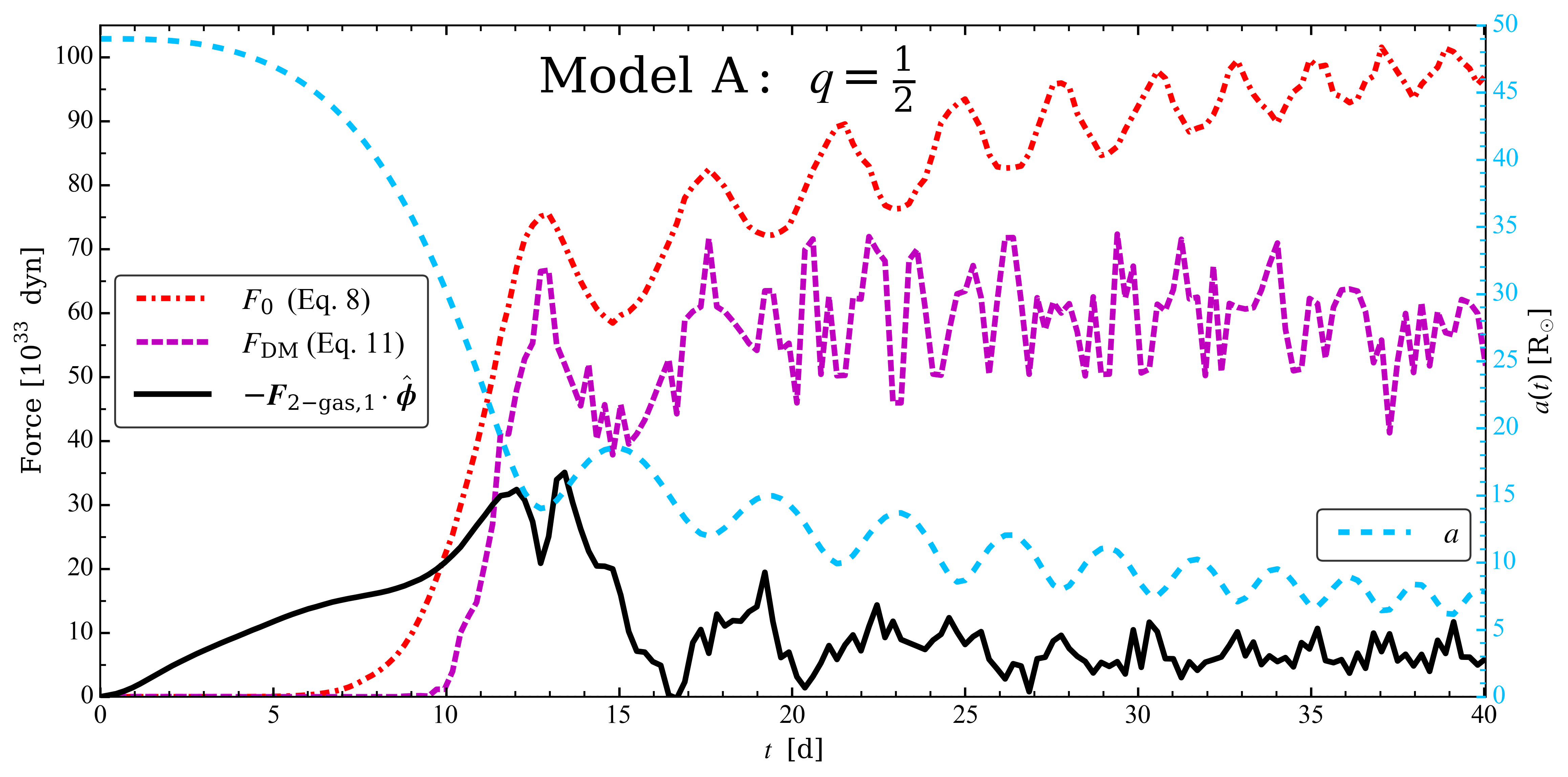}\\
  \includegraphics[width=0.9\textwidth,clip=true,trim= 0 40 0 0]{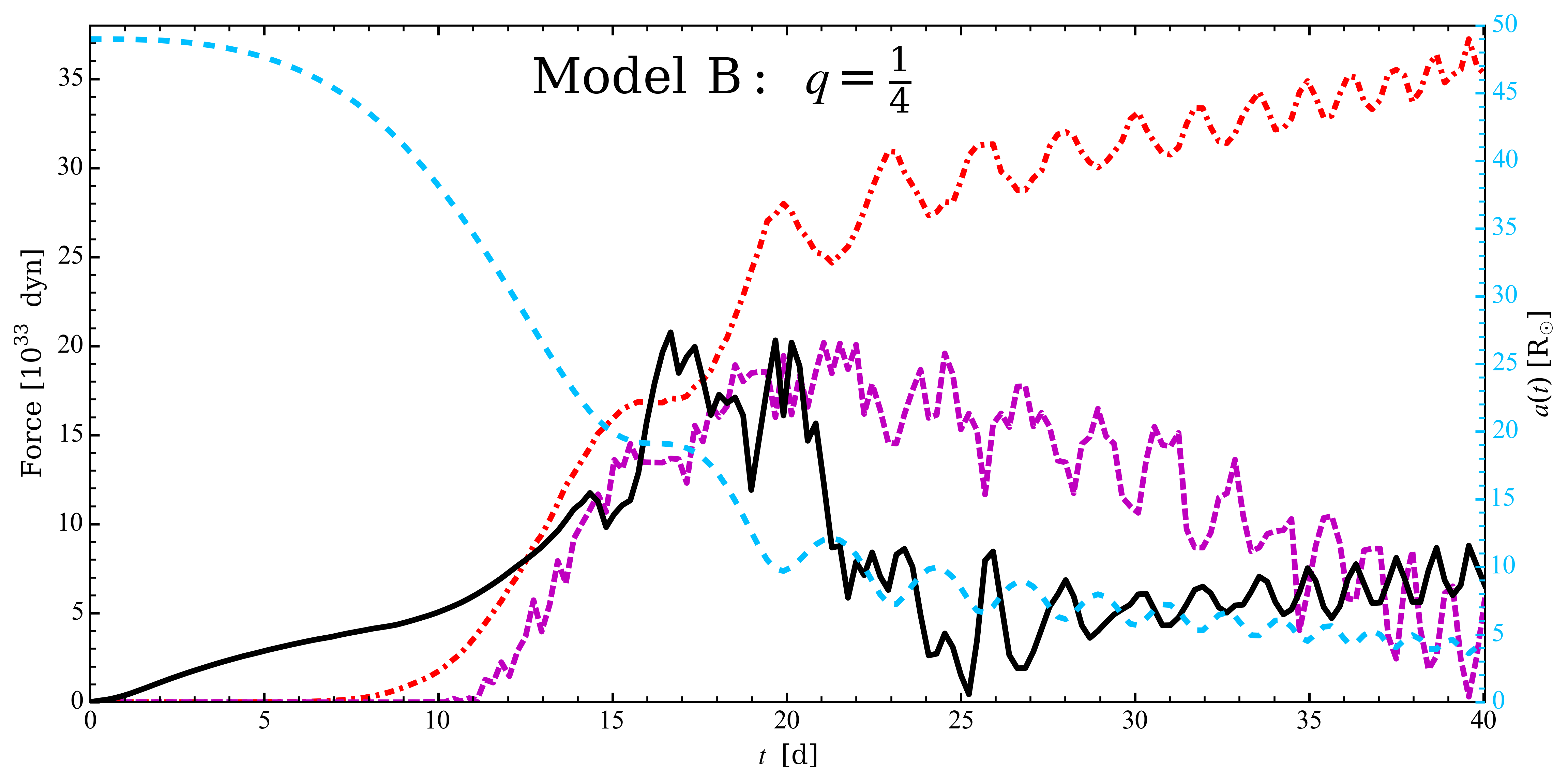}\\
  \includegraphics[width=0.89\textwidth,clip=true,trim= 0 0 4 0]{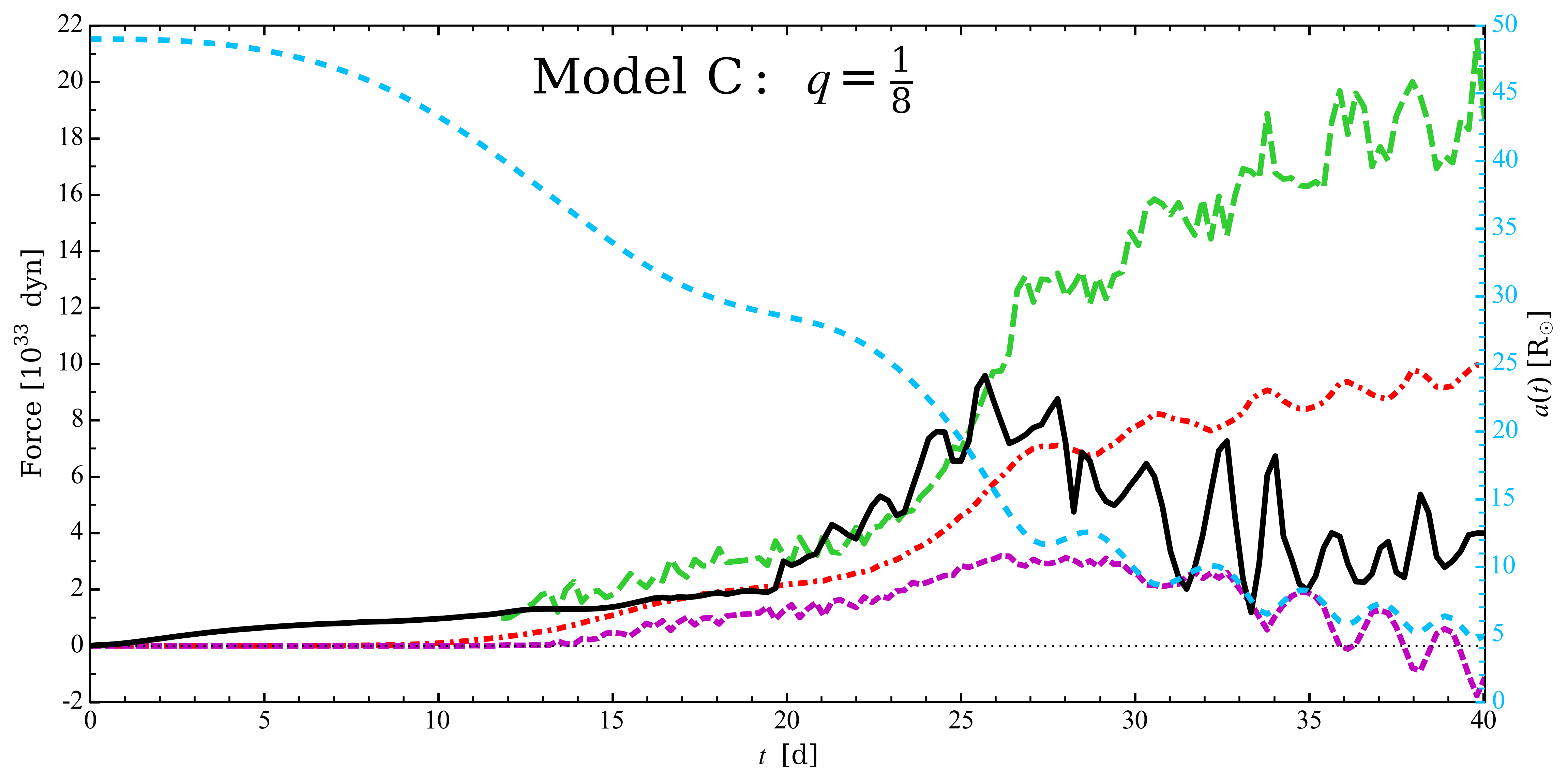}
  \caption{Azimuthal component of the net force on particle~2 due to the gas in the non-inertial rest frame of particle~1 
           (as in Fig.~\ref{fig:force2}), along with model predictions. 
           See text for explanations of the quantities plotted.
           The dashed green line in the bottom panel is from the fitting formula \eqref{F_M17},
           obtained from local 3D wind tunnel simulation results \citepalias{Macleod+17}. 
           This comparison is only carried out for Model~C,
           where $q\enc(t)$ is comparable with the value $q\enc=0.1$ used in their local simulations.
           \label{fig:f2theory}
          }            
\end{center}
\end{figure*}

\subsection{Estimate for Uniform Density}
The dynamical friction force can be estimated from Bondi-Hoyle-Lyttleton (BHL) theory
\citep{Hoyle+Lyttleton39,Bondi+Hoyle44,Bondi52}.
Here, gas approaching with impact parameter less than the accretion radius 
\begin{equation}
  \label{Ra}
  R_\mathrm{a}=\frac{2\Gn M}{c_\infty^2+v_\infty^2}
\end{equation}
accretes onto the star,
where $c_\infty$ and $v_\infty$ are the sound speed of the unperturbed envelope,
and its speed relative to the secondary.
The accretion rate can be estimated as $\dot{M}\sim\uppi R_\mathrm{a}^2\rho_\infty(c_\infty^2+v_\infty^2)^{1/2}$, 
where $\rho_\infty$ is the unperturbed density,
resulting in a  drag force  
\begin{equation}
  \label{Fd}
  F\sim\dot{M}v_\infty\ln\left(\frac{r_\mathrm{max}}{r_\mathrm{min}}\right)
   \sim\frac{4\uppi \Gn^2 M^2\rho_\infty v_\infty}{(c_\infty^2+v_\infty^2)^{3/2}}\ln\left(\frac{r_\mathrm{max}}{r_\mathrm{min}}\right).
\end{equation}
Typically, $r_\mathrm{max}$ is taken to be $R_\mathrm{a}$ and $r_\mathrm{min}$ as the radius of the star.
Equation (\ref{Fd}) was first derived by
\cite{Dokuchaev64} and survives among different estimates \citep{Edgar04} subjected to refinements from numerical studies, e.g. \citet{Shima+85}.
We neglect turbulence \citep{Krumholz+06}  which may be important in general.
We do consider the influence of a density gradient, as explained below.

To make contact with previous work, 
we plot the $\phi$-component of the drag force, as in Fig.~\ref{fig:force2},
but now with additional lines representing theoretical predictions or results from local simulations, 
in Fig.~\ref{fig:f2theory}.
The dash-dotted red line shows the quantity
\begin{equation}
  \label{F0}
  F\f\equiv \frac{4\uppi\Gn^2M\2^2\rho\f v\f}{(c\f^2+v\f^2)^{3/2}},
\end{equation}
where $v\f\equiv|\bfv\2-\bfv\1|\f$.
In our notation, quantities with a `0' subscript are computed from the initial envelope profile at radius $a(t)$,
with velocity 
computed assuming a circular orbit with primary mass equal to the mass interior to the orbit, $m\1(a)=M_\mathrm{1,c}+m\env(a)$.
We find that replacing $v\f$ by its actual value measured in the simulation $|\bfv\1-\bfv\2|$
 increases the amplitude of the oscillations in $F\f$ but otherwise the results are similar,
so we  opt to use the relative velocity computed from the initial profile.%
\footnote{Defining the Mach numbers $\Ma=|\bfv\2-\bfv\1|/c\f$ or $\Ma\f=v\f/c\f$,
these are found to be in the range $1.1<\Ma,\Ma\f<5.8$. 
Within this range, the values are larger for larger companion mass.
Both peak at the same time, at about $6$, $7$ and $9\da$ for Models~A, B and C, respectively, 
before leveling  near to the minimum as the simulation progresses.}

At early times, $F\f$ is effectively zero  due to the small $\rho\f$.
Subsequently, $F\f$ rises to be comparable to $\phi$-component of $-\bm{F}_\mathrm{2-gas,1}$
just before the first periastron passage,
before continuing to rise, in contrast to the $\phi$-component of $-\bm{F}_\mathrm{2-gas,1}$, 
which decreases and then levels off.
Hence, equation~\eqref{F0} overestimates the magnitude of the drag force at late times.
Qualitatively similar results were obtained by \citet{Staff+16b}.
However, the phase and amplitude of the variability seen in the $\phi$-component of $-\bm{F}_\mathrm{2-gas,1}$
at late times in Models~A and B is reproduced by this minimalist theoretical model.

\subsection{Estimate Including Density Gradient}
Using a more refined version of the theory that includes the logarithmic factor of equation~\eqref{Fd}
and accounts to some extent for gradients in the radial direction might be expected to produce better agreement.
Hence, for a more general estimate to the drag force, we multiply $F\f$ by $\ln(r_\mathrm{max}/r_\mathrm{min})$
and a correction factor \citet{Dodd+Mccrea52} (hereafter \citetalias{Dodd+Mccrea52}) $R_\mathrm{a,DM}^2/R_\mathrm{a,0}^2$, 
that accounts for a linear or at most quadratic density gradient in the correction to the accretion radius, computed by
\begin{equation}
  \label{R_DM}
  R_\mathrm{a,DM}= \frac{R_\mathrm{a,0}}{1+R_\mathrm{a,0}^2/(4H_\rho^2)},\\
\end{equation}
where 
\begin{equation}
  \label{Ra0}
  R_\mathrm{a,0}= \frac{2\Gn M\2}{c\f^2+v\f^2}
\end{equation}
and $H_\rho= -\rho\f/(\mathrm{d}\rho/\mathrm{d}r)\f$ is the scale height.
The modified force magnitude is then
\begin{equation}
  \label{F_DM}
  F_\mathrm{DM}= F\f\ln\left(\frac{r_\mathrm{max}}{r_\mathrm{min}}\right)\left(\frac{R_\mathrm{a,DM}}{R_\mathrm{a,0}}\right)^2.
\end{equation}
We adopt $r_\mathrm{max}=R_\mathrm{a,DM}$ and $r_\mathrm{min}= r\soft|_{t=0}=2.4\Rsun$.
The dashed purple line shows the resulting corrected estimate.%
\footnote{Radial variations in the density gradient of the initial envelope profile cause the noise. 
This variation is present in the 1D \textsc{mesa} solution, 
which was retained for our initial condition outside of $r=2.4\Rsun$.}

We see from Fig.~\ref{fig:f2theory} that despite some differences, 
the level of agreement between theory which includes the density gradient (dashed purple) 
and the simulation results (solid black) is overall comparable to that obtained using $F\f$ (dashed red).
The \citetalias{Dodd+Mccrea52} correction marginally improves agreement for Models~A and B,
but marginally worsens agreement for Model~C.\footnote{We also tried other variations, 
with only the $\ln(r_\mathrm{max}/r_\mathrm{min})$ factor or only the \citetalias{Dodd+Mccrea52} correction included, and found results that are generally similar to the cases plotted.}

\begin{figure*}
\begin{center}
  \includegraphics[width=0.8\textwidth,clip=true,trim= 0 40 0 0]{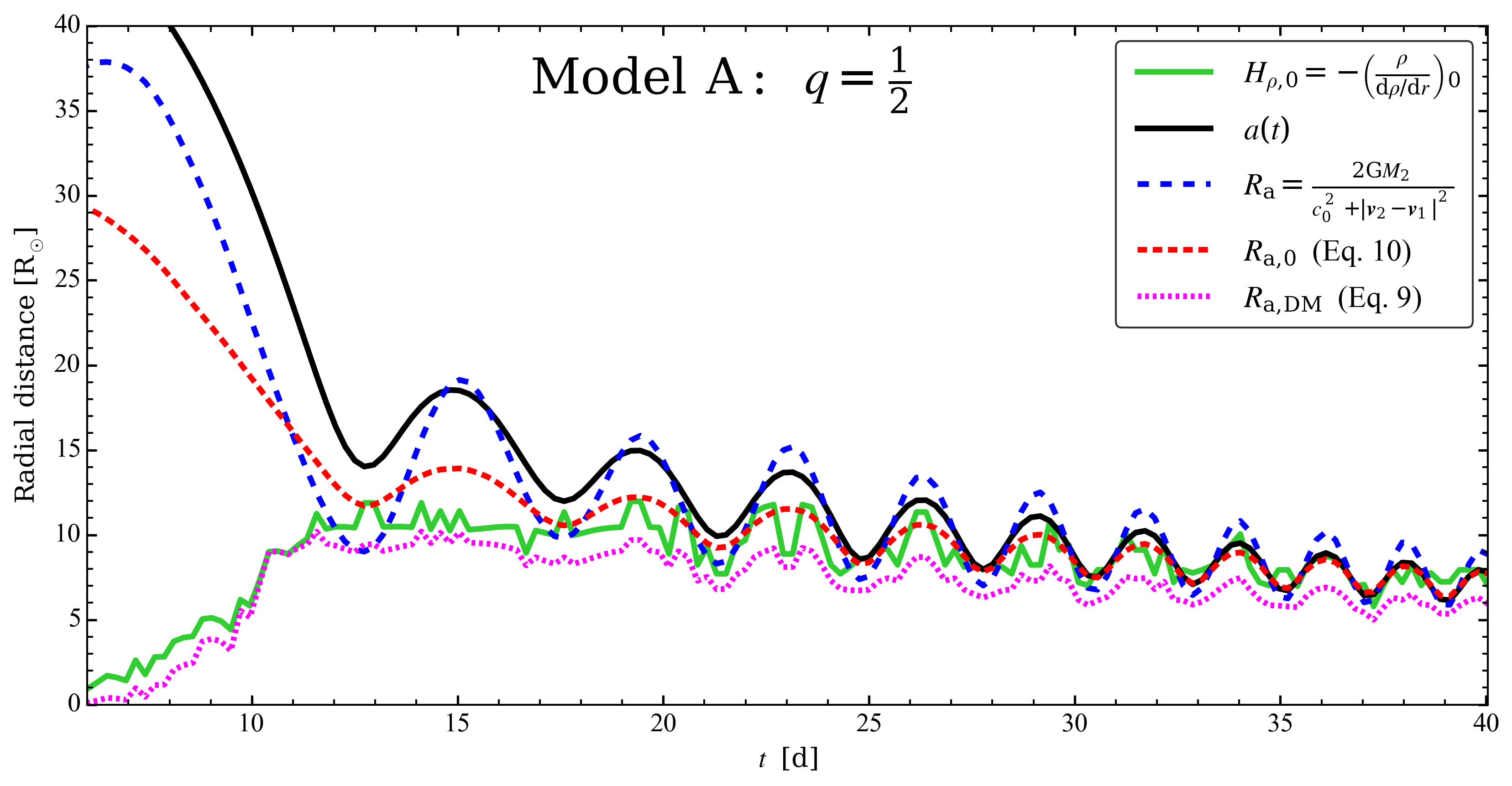}\\
  \includegraphics[width=0.8\textwidth,clip=true,trim= 0 40 0 0]{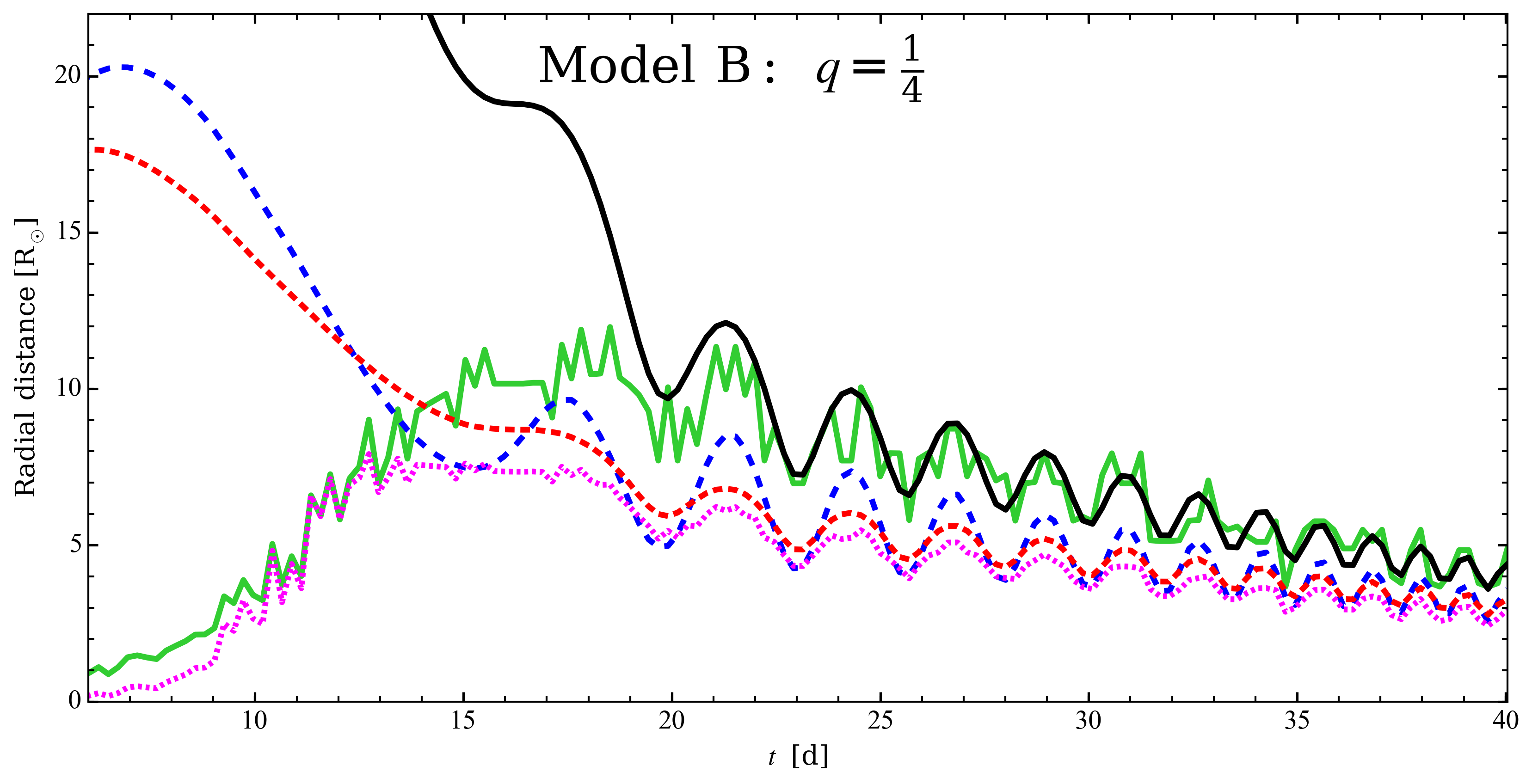}\\
  \includegraphics[width=0.8\textwidth,clip=true,trim= 0  0 0 0]{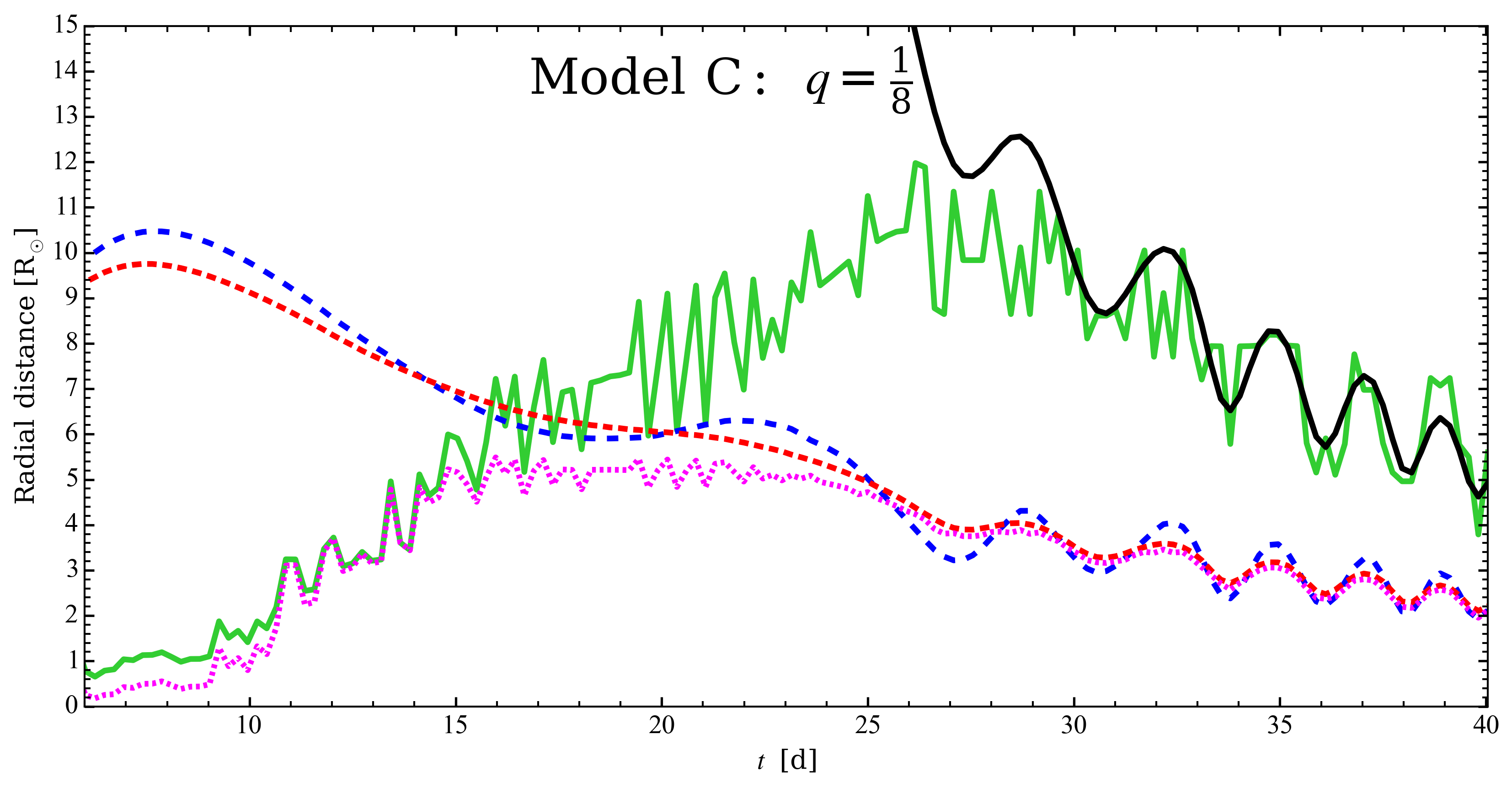}
  \caption{Comparison between various relevant length scales, plotted against time. 
           \label{fig:Ra}
          }            
\end{center}
\end{figure*}

\subsection{Theory works best at intermediate times}
We do not expect good agreement between simulation and theory at early times because of three interrelated factors: 
(i) tidally drawn envelope material increases the density near particle~2 beyond ambient and outer envelope layer values,
(ii) density scale heights are initially small compared to the accretion radius
and (iii) the initial condition of a secondary placed just outside the spherically symmetric primary at $t=0$
is not fully realistic.
Point (ii) is seen by comparing solid black and green lines in Fig.~\ref{fig:Ra}, 
where we plot the various length scales as a function of time, for each simulation.
The \citetalias{Dodd+Mccrea52} correction in principle helps to account for (ii)
but considers only the lowest order effect of the density gradient.

At late times, we also expect poor agreement.
In Model~A (q=1/2) $R\acc\sim a$ shortly after the first periastron passage, 
as seen in Fig.~\ref{fig:Ra} (compare blue and red dashed lines with black solid line),
so we do not expect good agreement.
For Model~B $(q=1/4)$, $R\acc$ remains marginally smaller than $a$,
while for Model~C (q=1/8), $R\acc\sim 0.5 a$  by the end of the simulation.
Theoretical predictions for late times improve slightly as $q$ decreases to $1/8$, but not dramatically.

At intermediate times, when $R\acc\ll a$ and $H_{\rho,0}\gtrsim R\acc$,
we expect and find agreement to be much better.
If we use $|\bfv\2-\bfv\1|$ measured directly from the simulation to compute $R\acc$ (dashed blue in Fig.~\ref{fig:Ra}),
then the time range when $R\acc(t)< a(t)$ at the previous periastron passage becomes $12\da\lesssim t\lesssim 14\da$ for Model~A, 
$13\da\lesssim t\lesssim 24\da$ for Model~B, and $t\gtrsim16\da$ for Model~C.
BHL/DM theory  approximates  the numerical results reasonably well in these time ranges.
In particular, during the broad force peak, theory (equation~\ref{F0} or \ref{F_DM})
correctly predicts the force to within a factor of $\sim2$ for all models.

\subsection{Improved theory is needed for late times}\label{sec:O99}

\citet{Reichardt+19} suggested that a reduction in the relative velocity between the particles and gas at late times 
in their simulation with $q=0.68$ might help explain the reduction of the drag force. 
Might replacing  our initial values of the gas density, velocity with respect to particle~2, and sound speed 
with those measured directly from the simulation help reconcile theory and simulation at late times? 

From the right column of Fig.~\ref{fig:2D} (see Sec.~\ref{sec:2D} for details), 
we can estimate the parameters $\rho_\infty$, $v_\infty$ and $c_\infty$ at $t=22.0\da$. 
Since the orbital separation is small, one option is to choose a location in the vicinity
of particle~2, namely in the region below and to the left of particle~2 and within a few $\!\Rsun$ of particle~2 in the plots.
The velocity vectors show that some of this gas is being sling-shotted around particle~2 
and this produces a drag force.
Here, $\rho_\infty\sim(5$--$10)\rho\f$, $v_\infty\sim\tfrac{1}{2}v\f$, and $c_\infty\sim2c\f$.
This leads to a force $F_\infty\sim(1.9$--$3.8)F\f$, using the analogue of equation~\eqref{F0},
so the predicted force is even larger than before, instead of smaller, as needed.
Perhaps more reasonable is to choose a location farther away from particle~2
on the circle centred on particle~1 with radius equal to $a$,
near coordinates $(-10,-5)$ in the right column of Fig.~\ref{fig:2D}.
In this vicinity, we find $\rho_\infty\sim3\rho\f$, $v_\infty\sim0.3v\f$, and $c_\infty\sim c\f$,
giving $F_\infty\sim 4.9F\f$.
Pushing the values as far as seems reasonable to obtain a smaller estimate for the force, 
we could instead choose $\rho_\infty\sim\rho\f$, $v_\infty\sim\tfrac{1}{2}v\f$ and $c_\infty\sim2c\f$,
which leads to $F_\infty\sim 0.38F\f$.
This value is still not small enough to explain the factor of $\sim10$ between the solid black and dash-dotted red lines 
in the top panel of Fig.~\ref{fig:f2theory} at $t=22.0\da$.

A second possibility is to note that 
the conventional theoretical drag force formula changes  as $\Ma_\infty$ becomes small \citep{Ostriker99}.  
Could this account for the small values seen at late times, as suggested by \citet{Staff+16b} for their simulations?
In the subsonic regime \citet{Ostriker99} derives the solution 
\begin{equation}
  \label{F_O99}
  F_\mathrm{O99} = -I\frac{4\uppi\Gn^2M^2\rho_\infty}{v_\infty^2},
\end{equation}
where
\begin{equation}
  I= \frac{1}{2}\ln\left(\frac{1+\Ma_\infty}{1-\Ma_\infty}\right)-\Ma_\infty.
\end{equation}
This leads to the factor $[(1+\Ma_\infty^2)^{3/2}/\Ma_\infty^3]I\approx0.48$ multiplying our above estimate of the force
if $v_\infty=\tfrac{1}{2}v\f$ and $c_\infty=2c\f$, 
or $\approx0.56$ if $V_\infty=0.3v\f$ and $c_\infty=c\f$ are used instead.
Thus, we obtain the new estimates $F_\infty\sim(1$--$2)F\f$, $3F\f$ and $0.18F\f$ for the three cases described above. 
Even the smallest of these is about a factor of two too large to explain the force measured in the simulation.
We obtain better agreement at $t=40\da$.
However, it is not clear whether equation~\eqref{F_O99} is even applicable in the present context.

Thus, such theoretical estimates 
are not well-motivated for late times owing to the small inter-particle separation, 
are sensitive to arbitrary choices, 
and cannot reproduce the force measured from the simulation. 

Another possibility is to note that at late times the particles accrete their own quasi-static, 
quasi-spherical ``bulges'' of gas \citepalias{Chamandy+18}.
The bulge around particle~2 is mainly pressure supported but partially rotation supported, 
and has size $r\bulge\sim2r\soft=2.4\Rsun$. 
Although particles interact with gas only through gravity, and thus can only experience dynamical drag,
the composite particle-bulge ``system'' in addition 
experiences a \textit{hydrodynamic} drag as it moves through the surrounding envelope gas.
The hydrodynamic drag force can be estimated as $F\hyd\sim\rho v^2 \uppi r\bulge^2$,
where $\rho v^2$ is the ram pressure exerted by the gas encountered by the bulge around particle~2 as it orbits.
For Model~A at $t=22.0\da$, using the above estimate $\rho\sim(5$--$10)\rho\f$ and $v\sim 0.5v\f$.
gives $F\hyd\sim(0.5$--$1)\times10^{34}\dyn$, which is just the order of magnitude needed.
However, this formula underestimates the drag force at $t=40\da$, and hence is inadequate.

We leave further exploration of the force at late times for future study. 

\section{Evolution of flow properties}\label{sec:2D}
\begin{figure*}
  \includegraphics[height=38.5mm,clip=true,trim= 120 165 225 180]{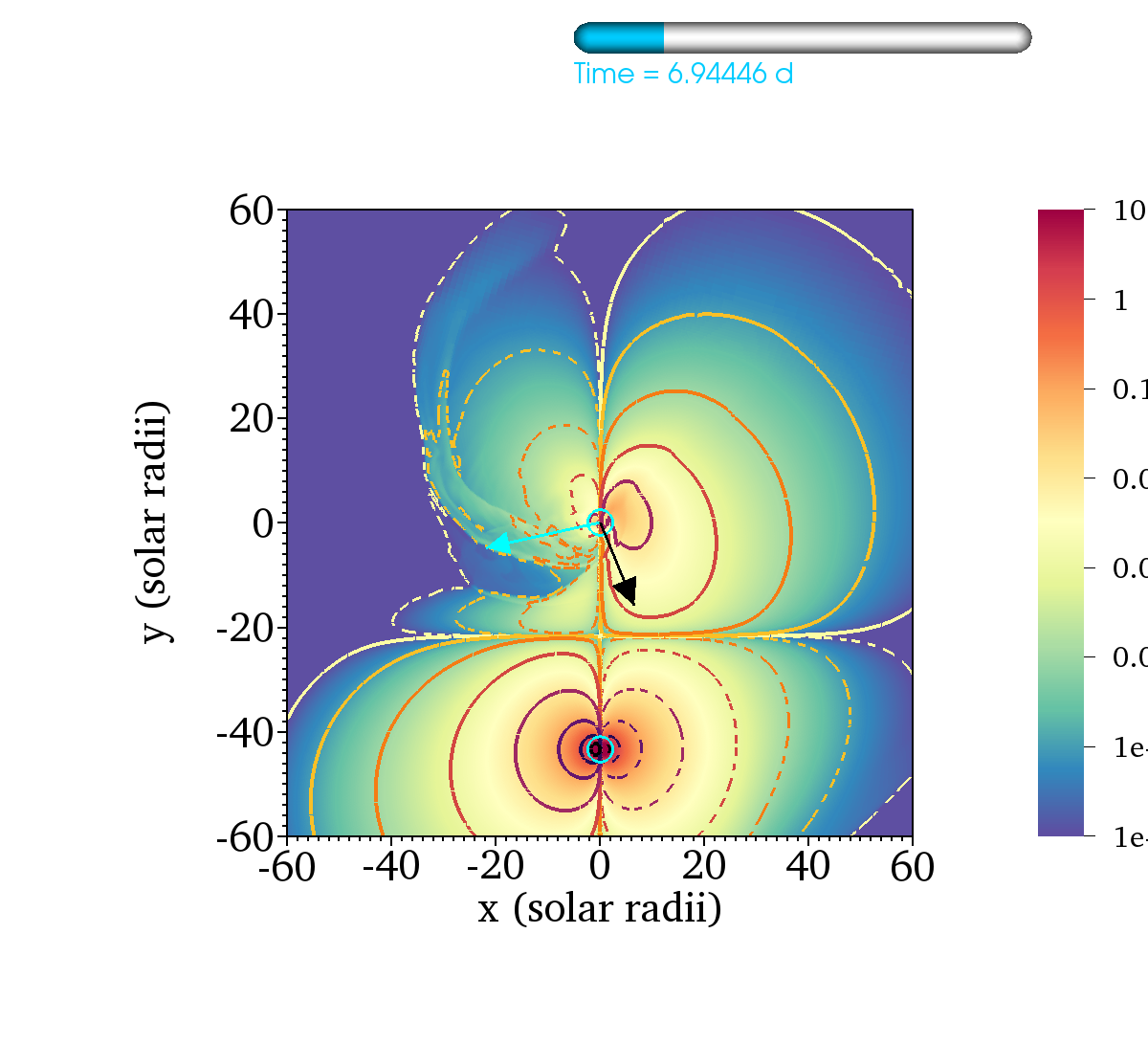}
  \includegraphics[height=38.5mm,clip=true,trim= 200 165 225 180]{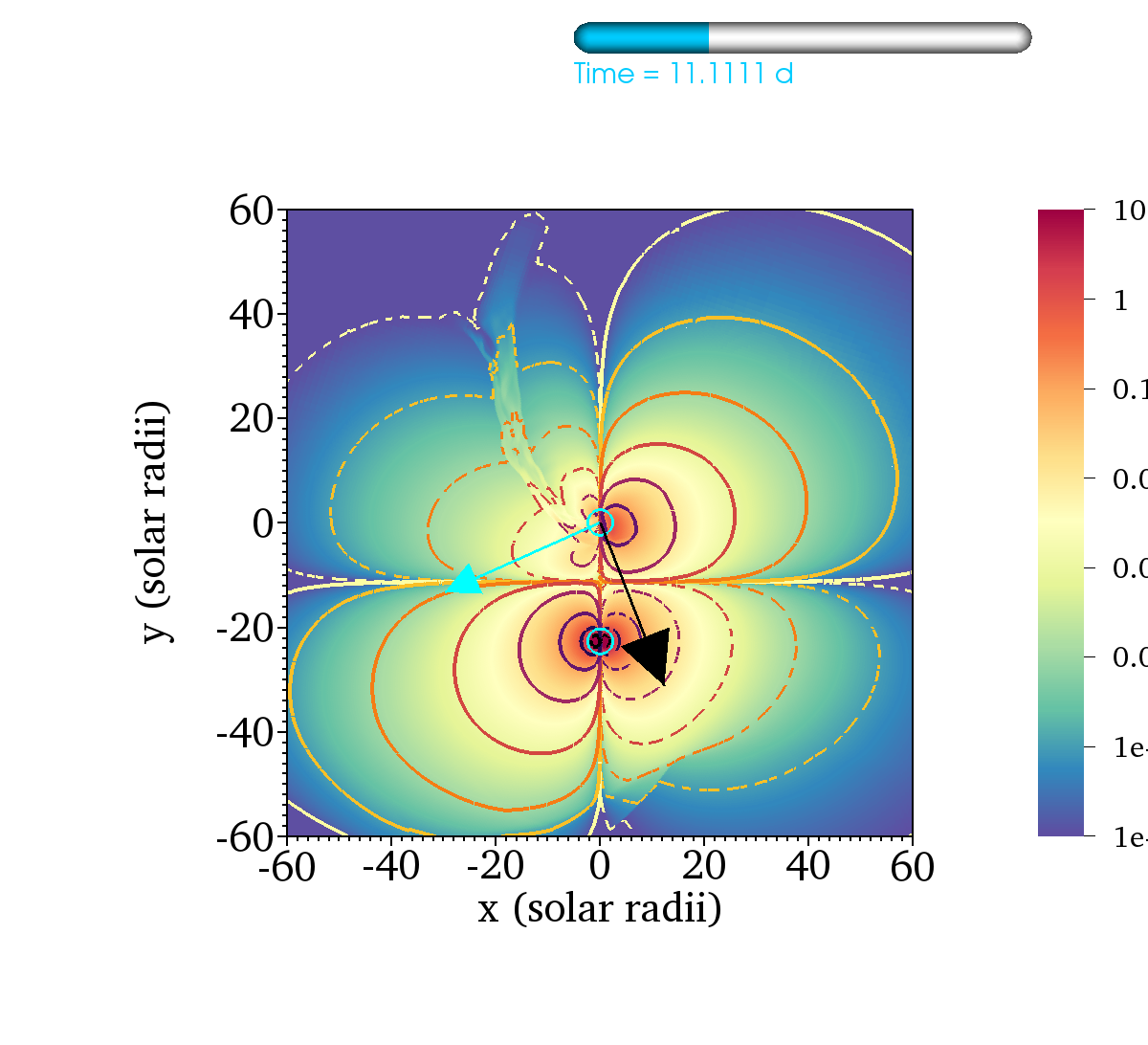}
  \includegraphics[height=38.5mm,clip=true,trim= 200 165 225 180]{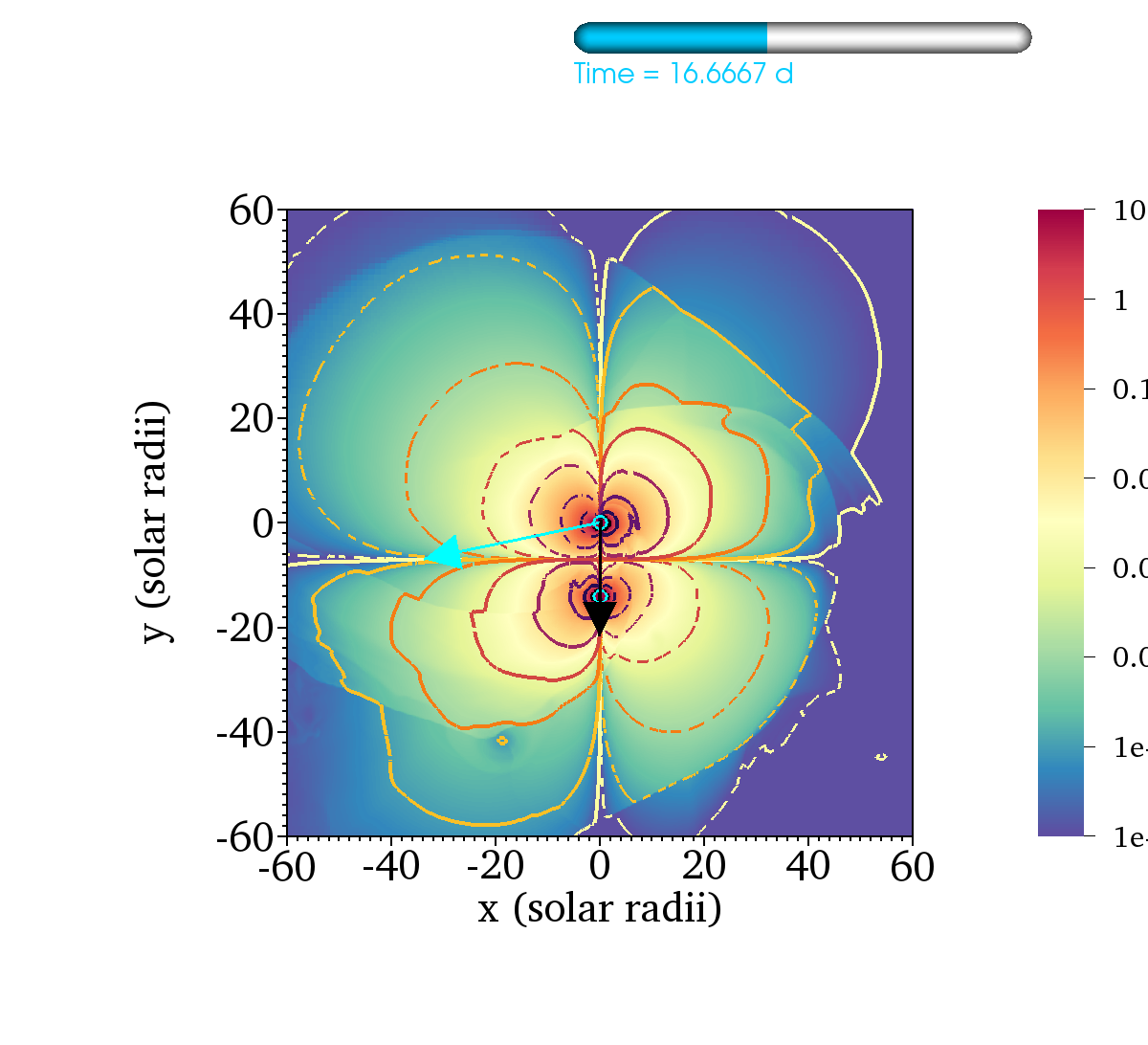}
  \includegraphics[height=38.5mm,clip=true,trim= 200 165  20 180]{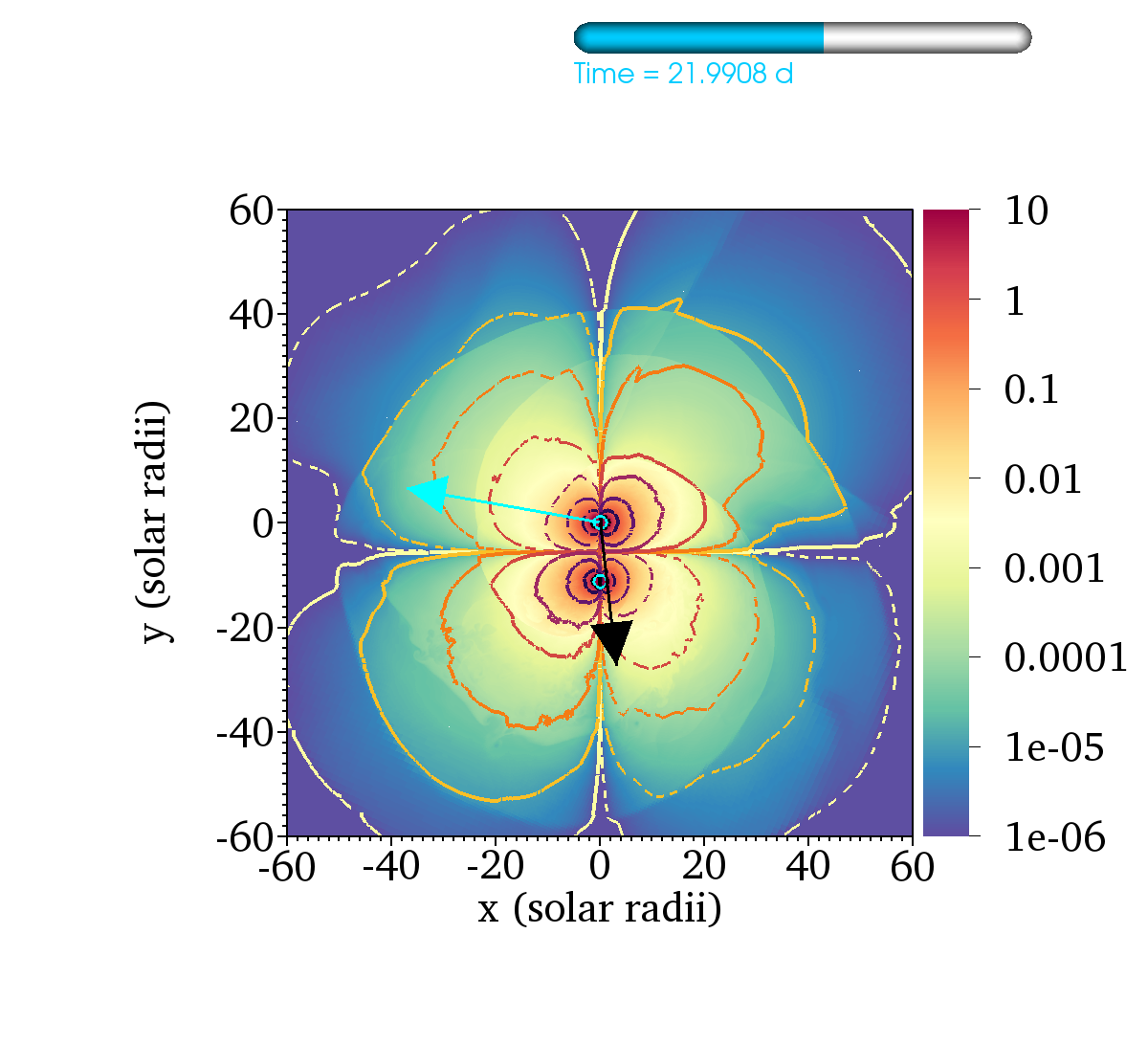}\\
  \includegraphics[height=38.5mm,clip=true,trim= 120 165 225 180]{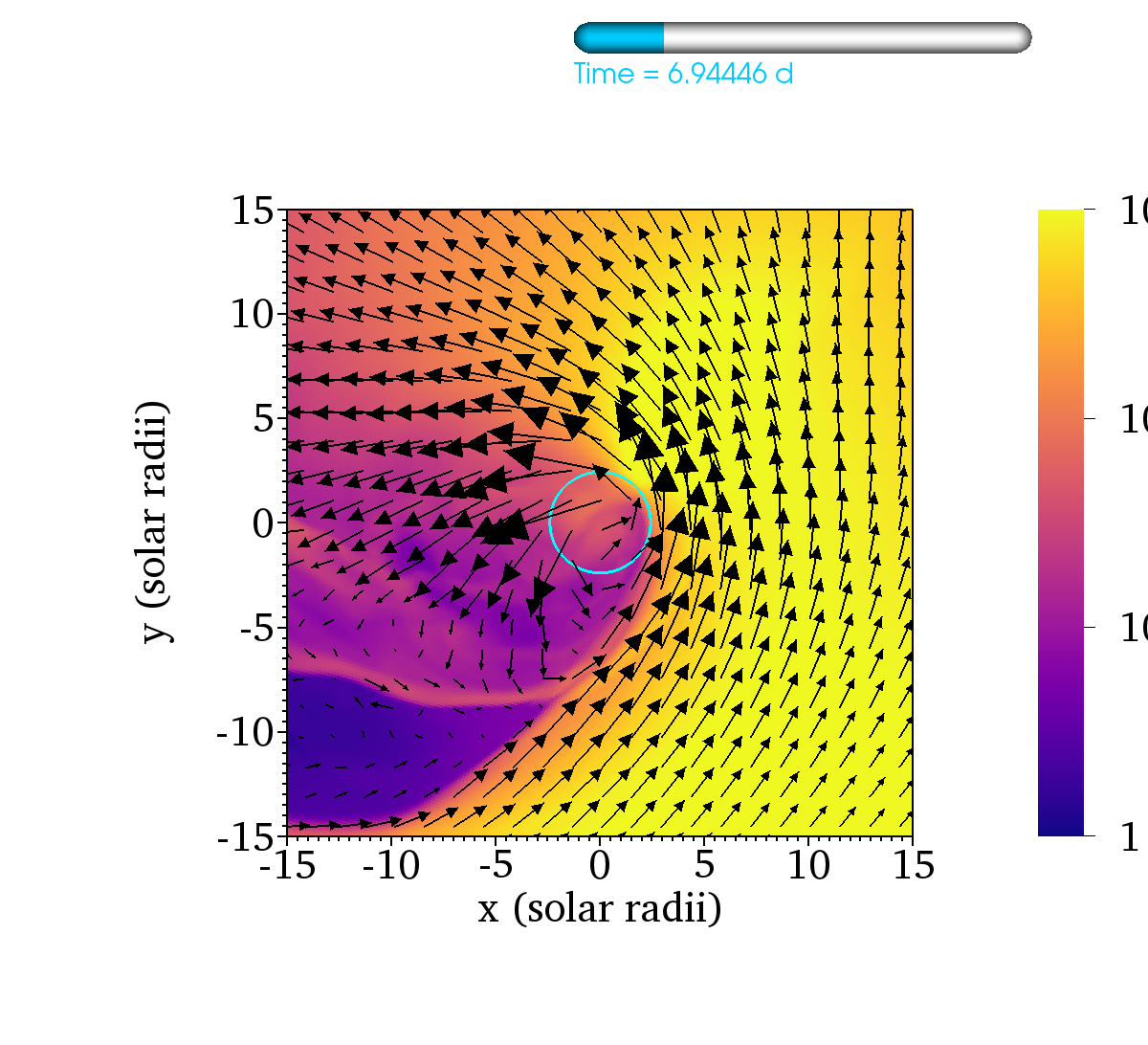}
  \includegraphics[height=38.5mm,clip=true,trim= 200 165 225 180]{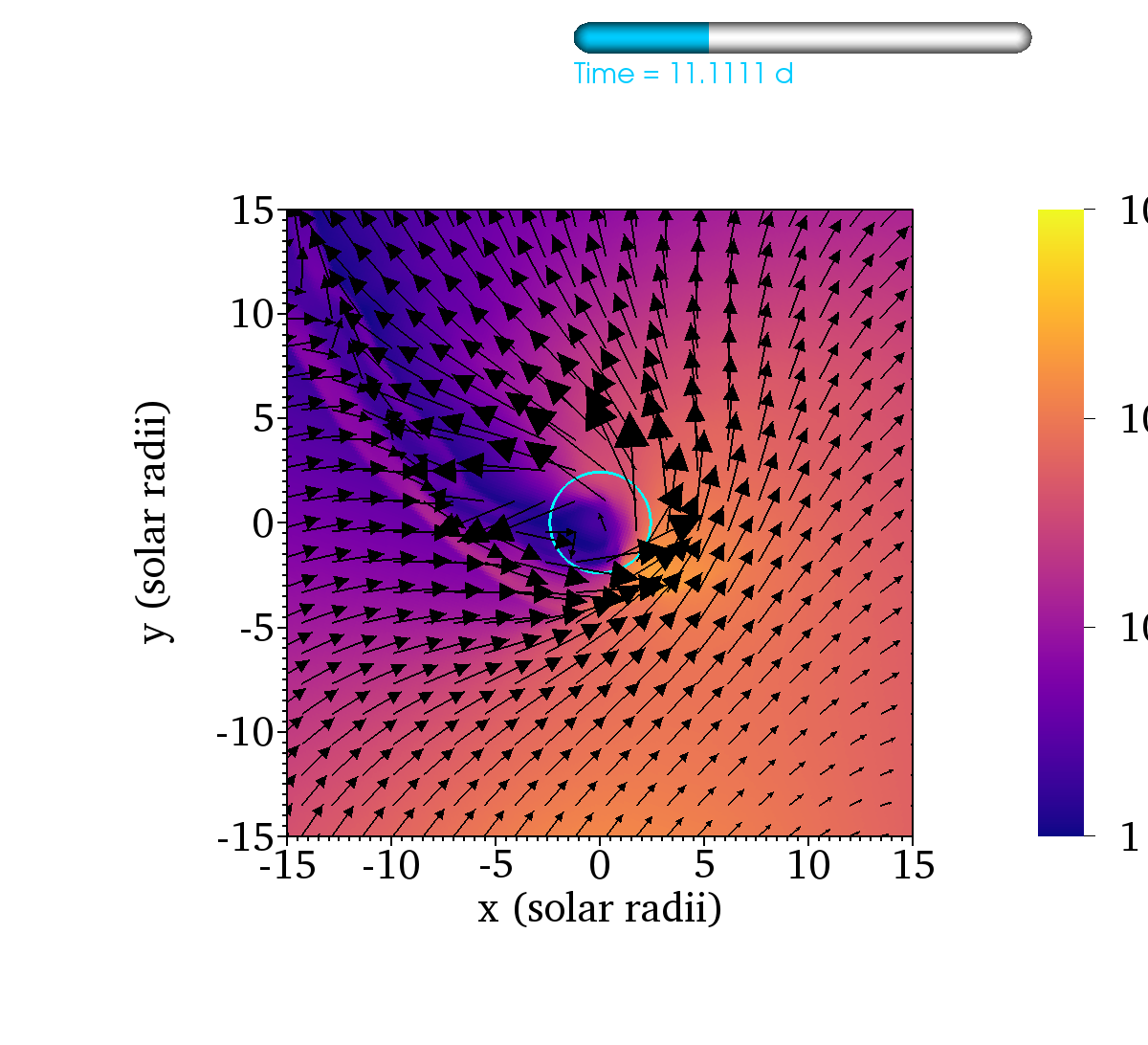}
  \includegraphics[height=38.5mm,clip=true,trim= 200 165 225 180]{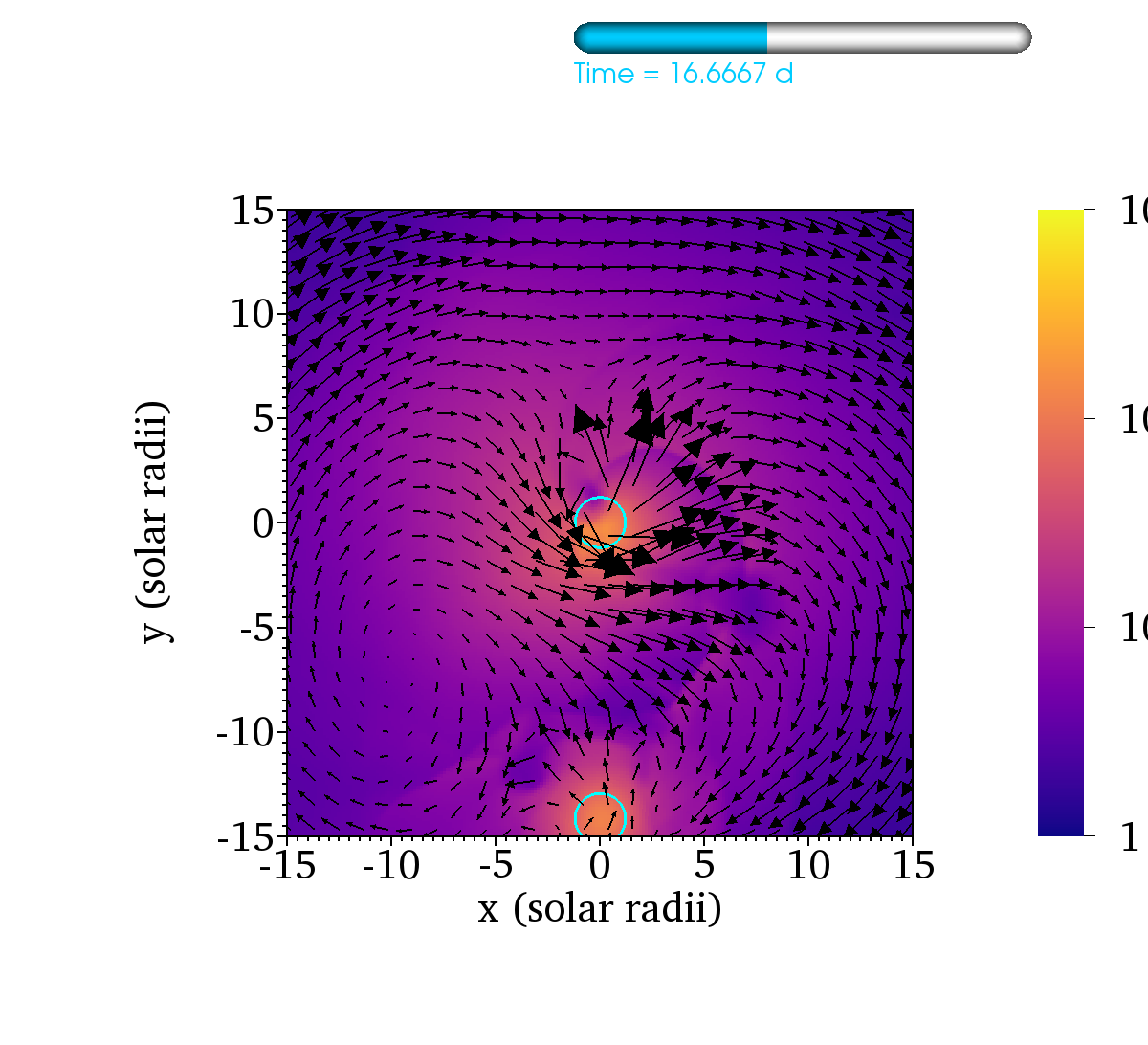}
  \includegraphics[height=38.5mm,clip=true,trim= 200 165  20 180]{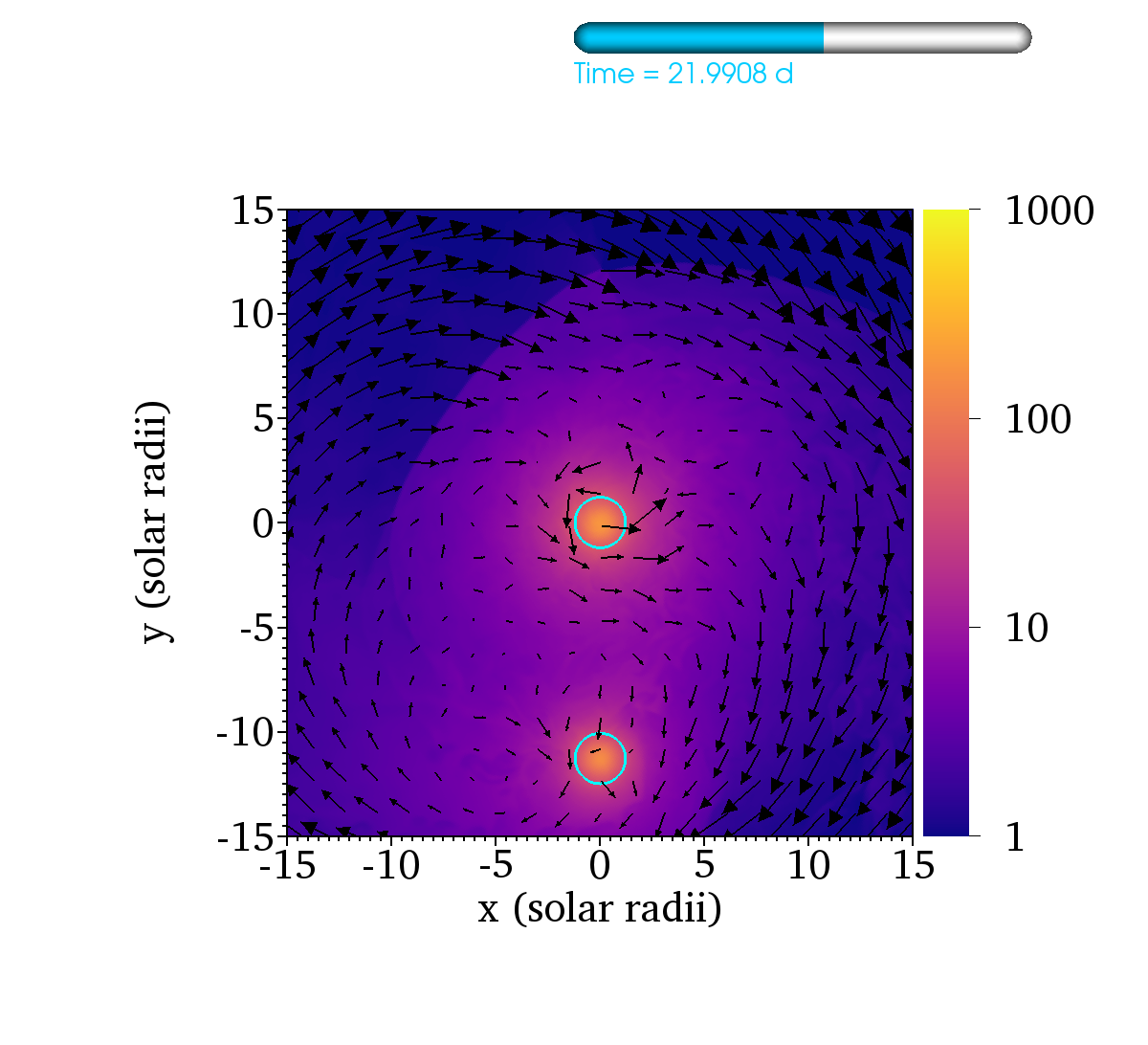}\\
  \includegraphics[height=38.5mm,clip=true,trim= 120 165 225 180]{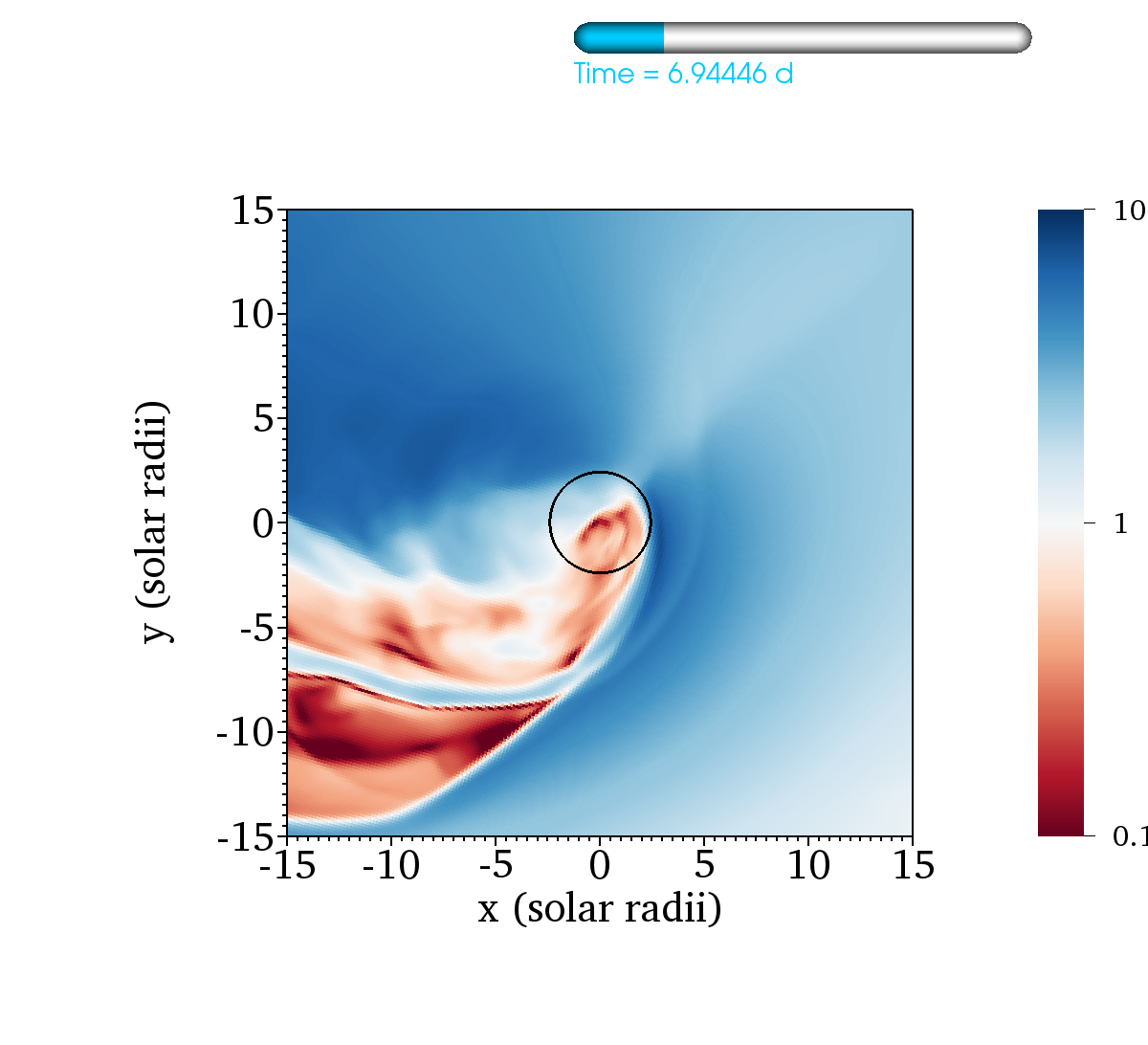}
  \includegraphics[height=38.5mm,clip=true,trim= 200 165 225 180]{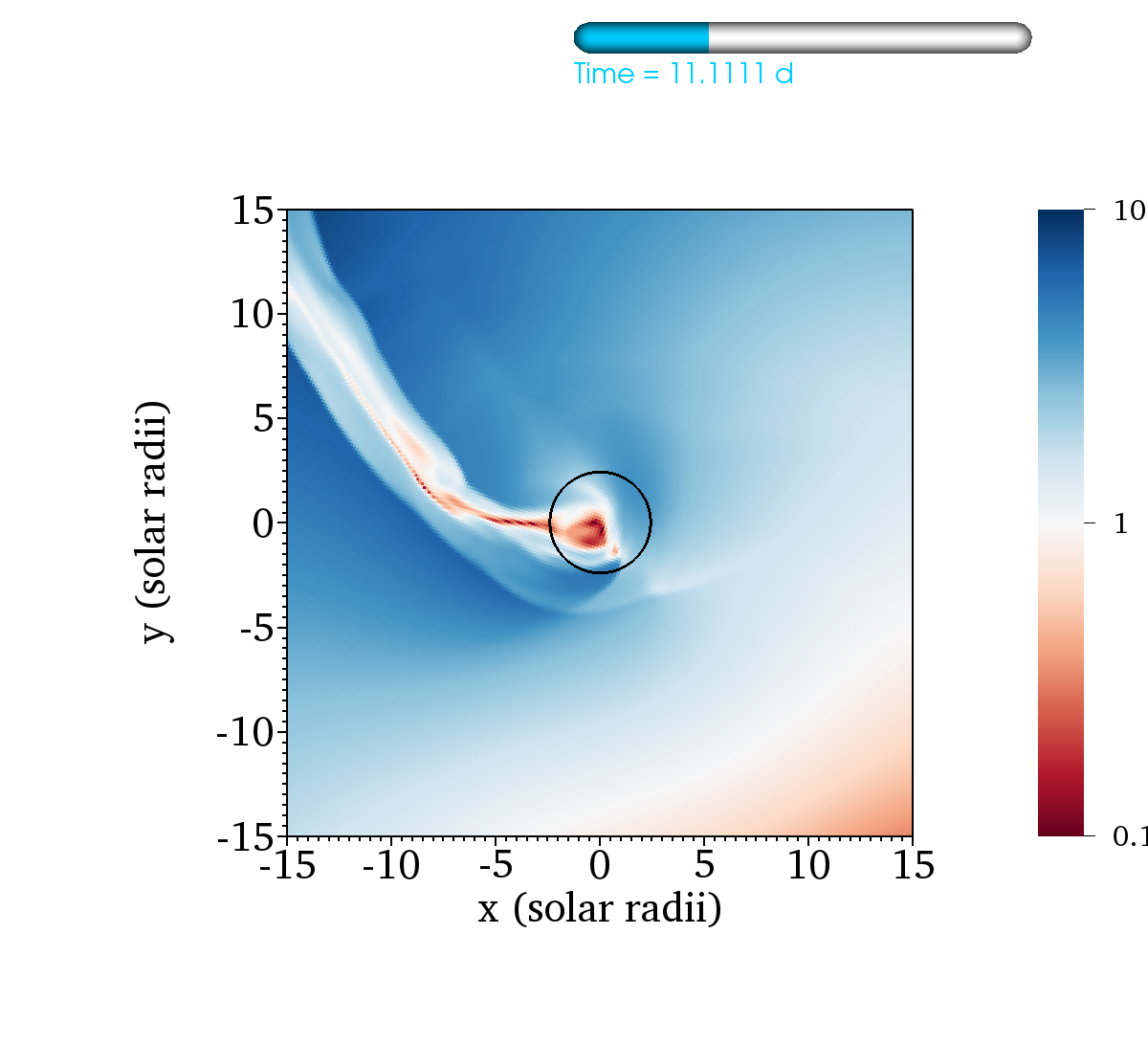}
  \includegraphics[height=38.5mm,clip=true,trim= 200 165 225 180]{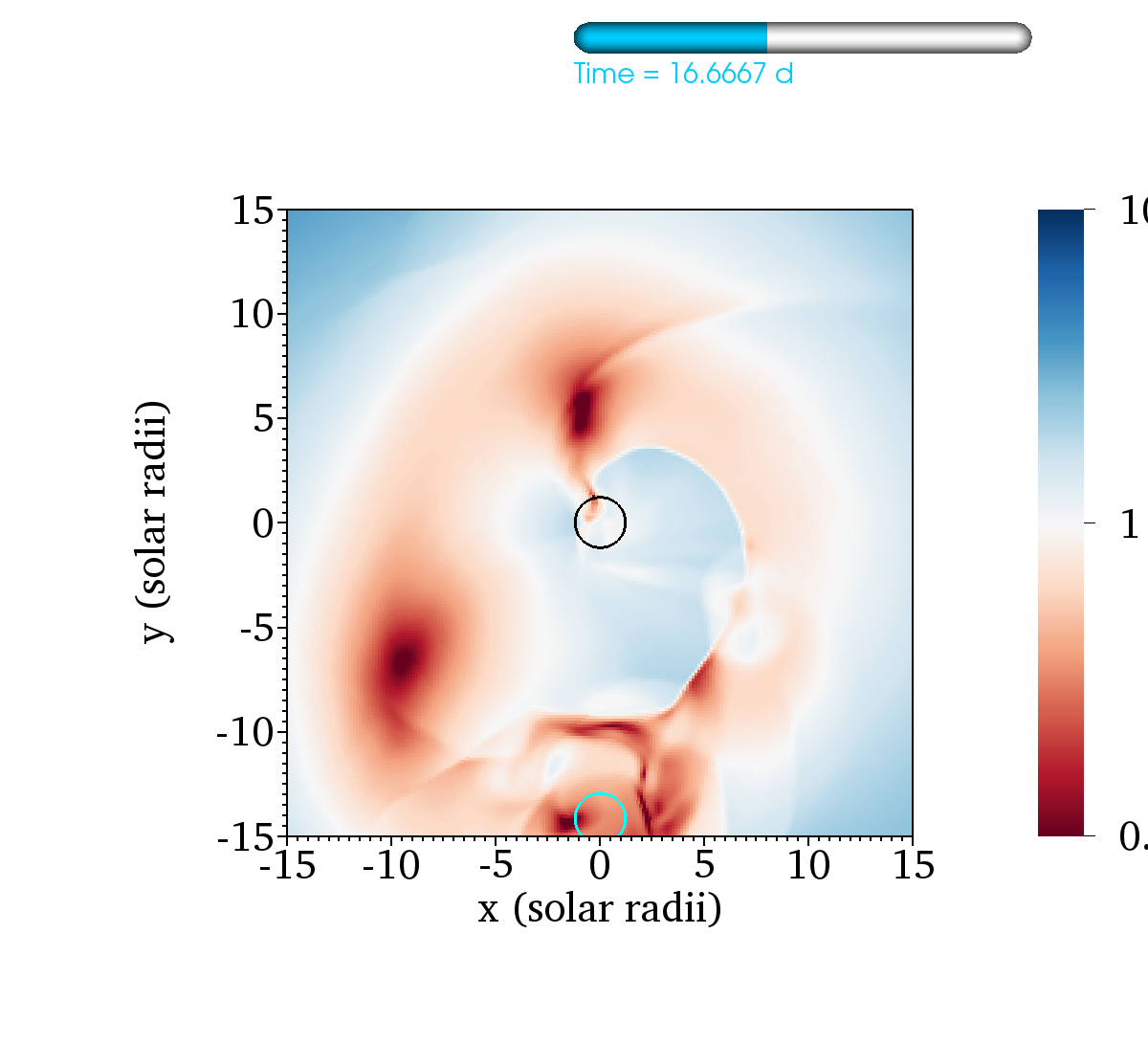}
  \includegraphics[height=38.5mm,clip=true,trim= 200 165  20 180]{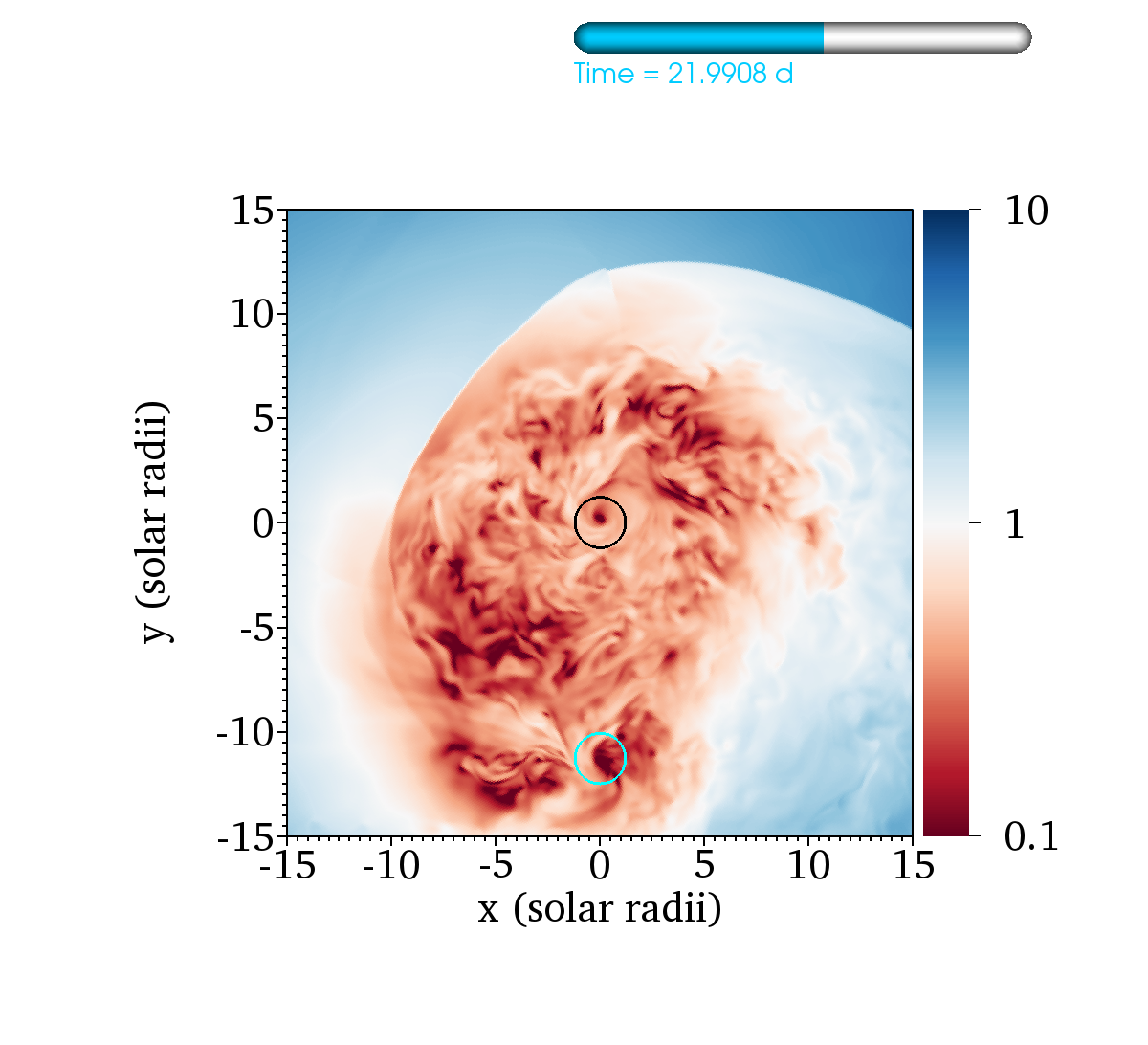}\\
  \includegraphics[height=38.5mm,clip=true,trim= 120 165 225 180]{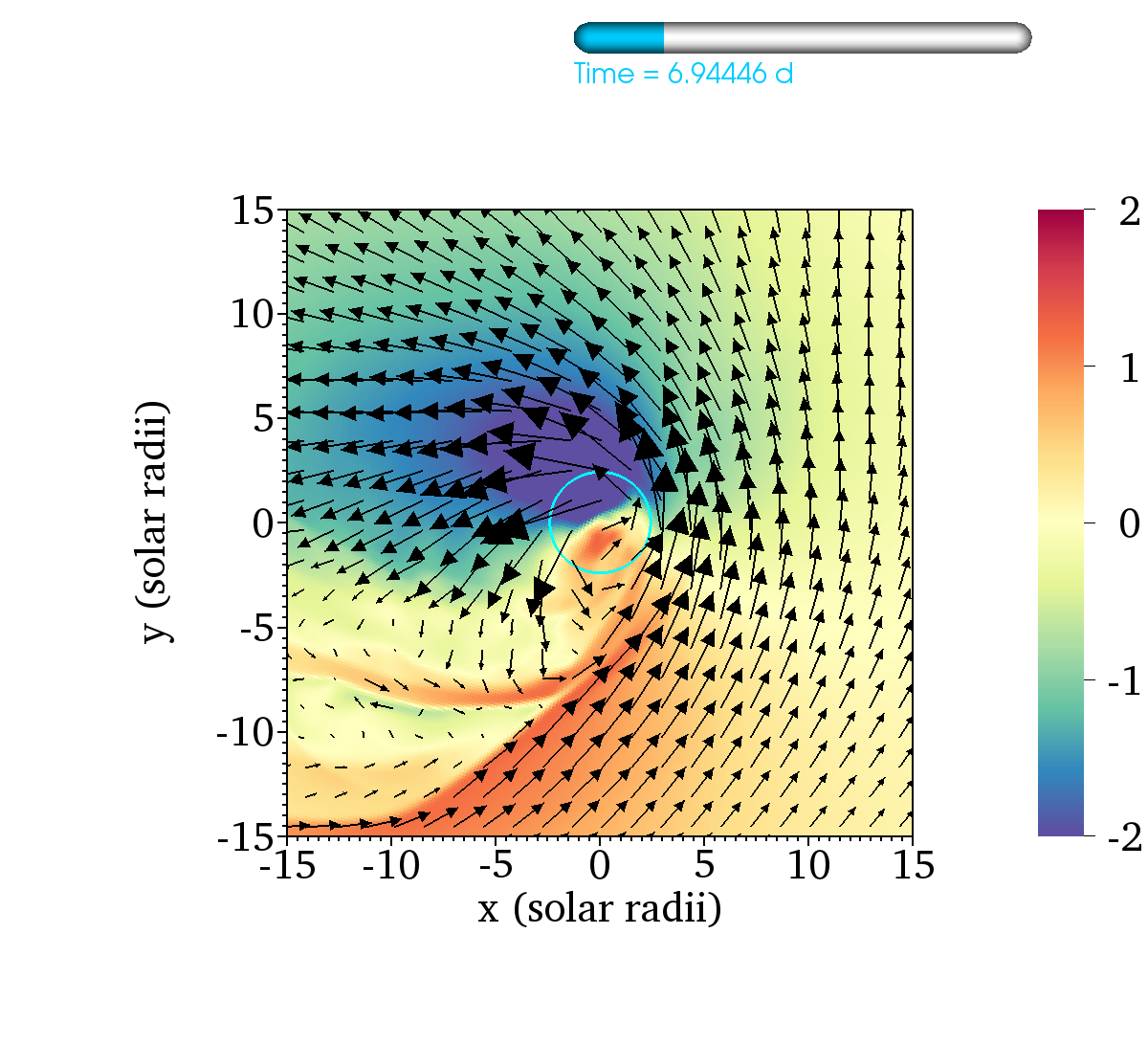}
  \includegraphics[height=38.5mm,clip=true,trim= 200 165 225 180]{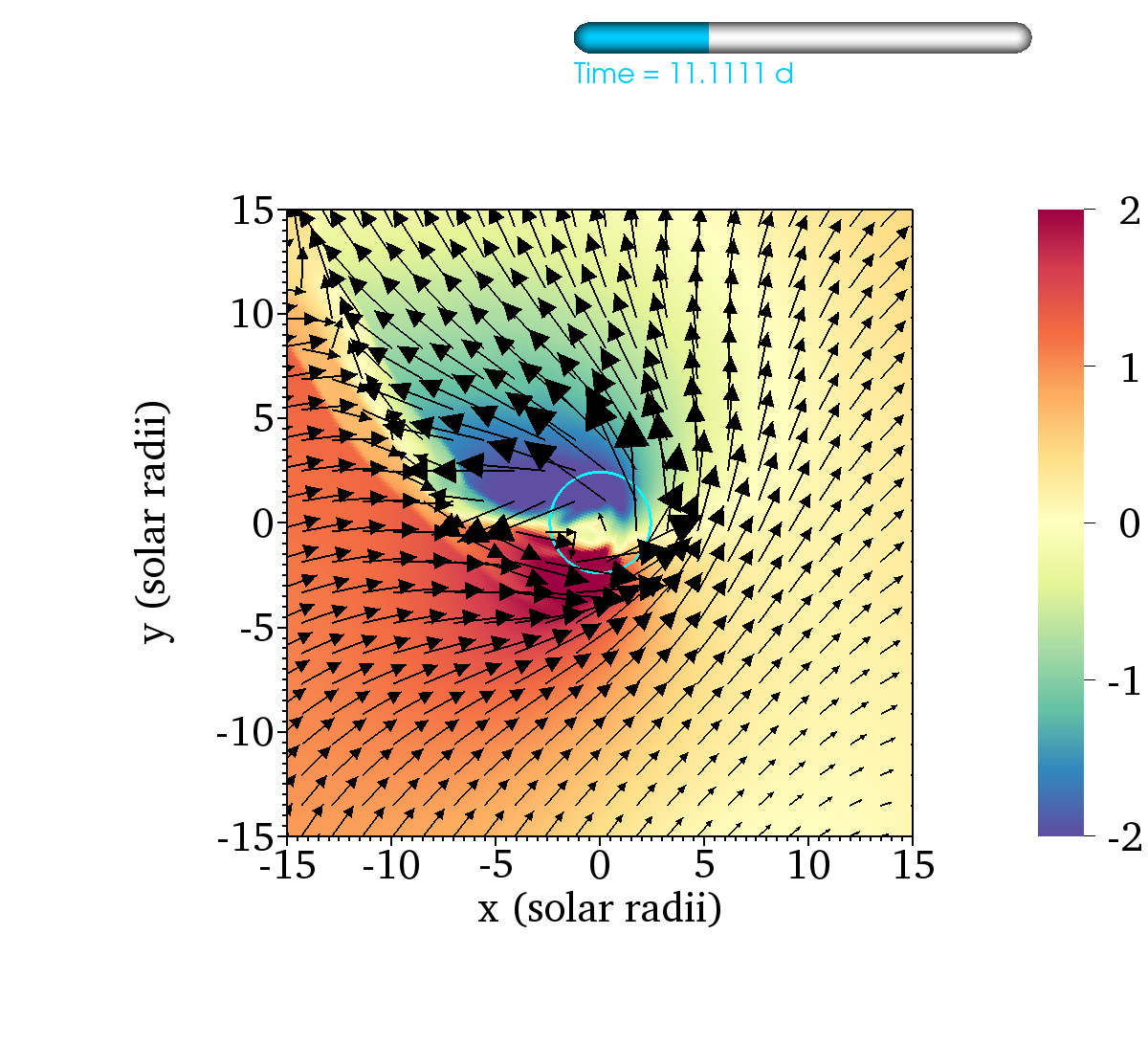}
  \includegraphics[height=38.5mm,clip=true,trim= 200 165 225 180]{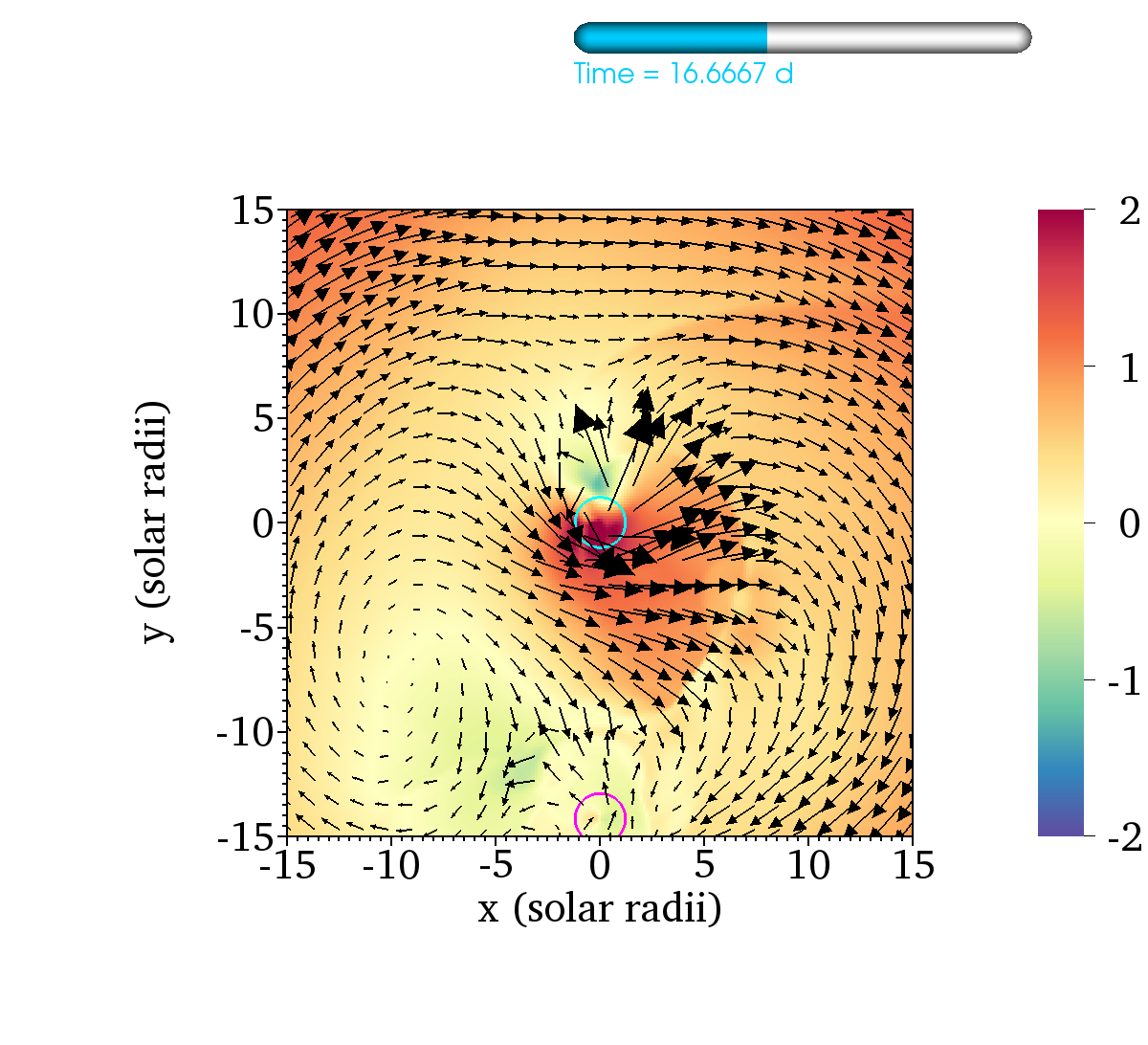}
  \includegraphics[height=38.5mm,clip=true,trim= 200 165  20 180]{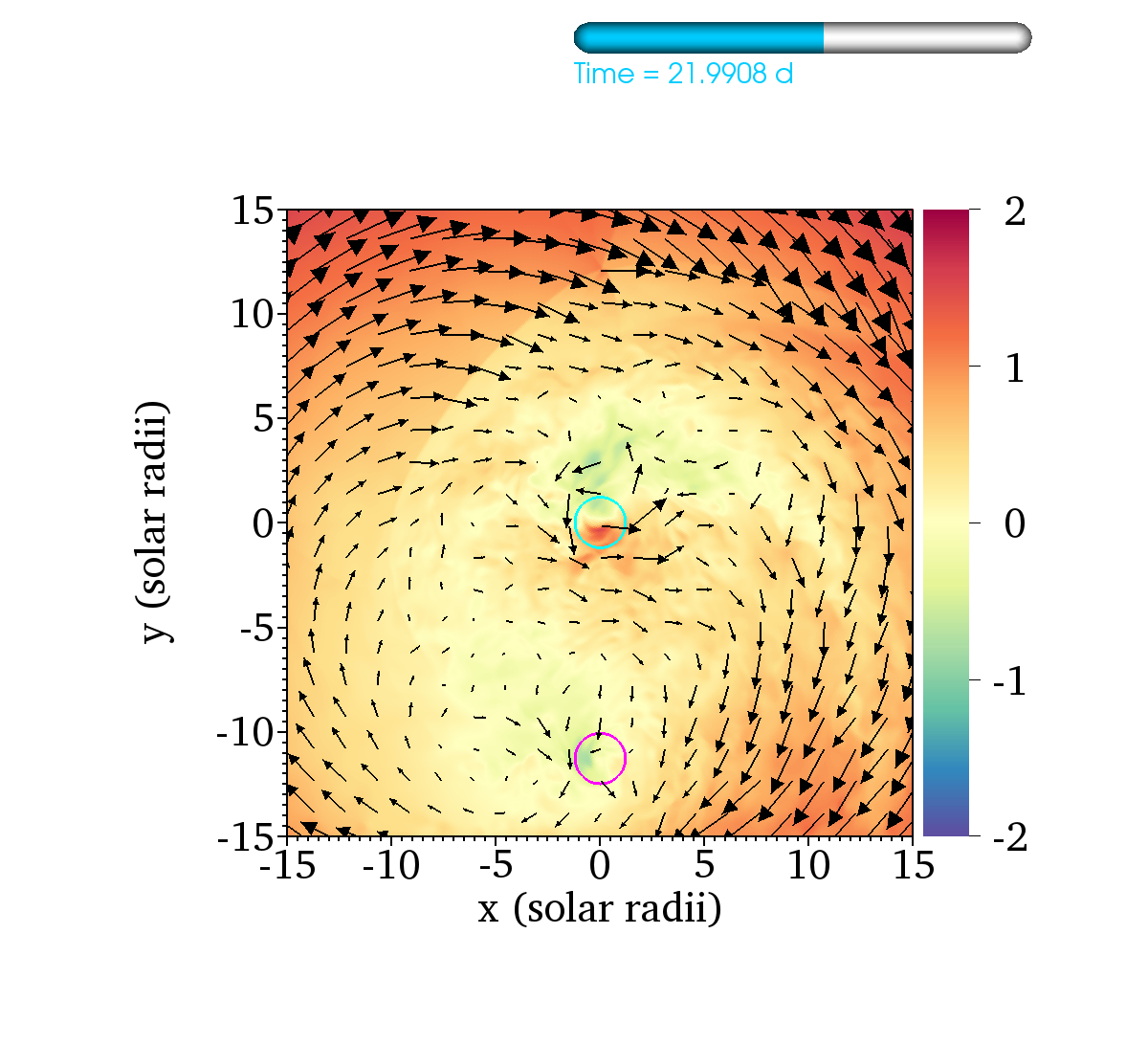}\\
  \includegraphics[height=40.55mm,clip=true,trim= 120 125 225 180]{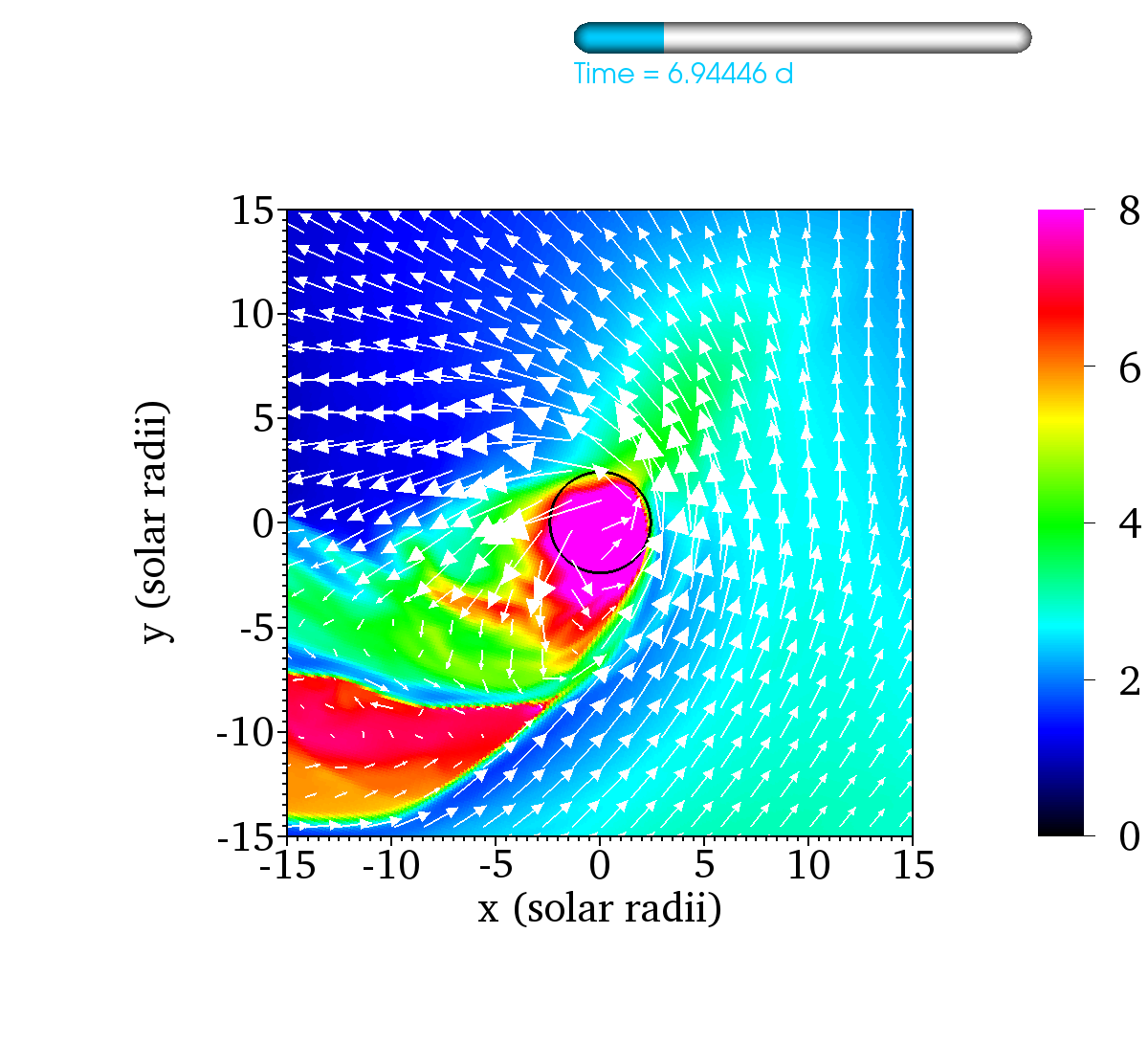}
  \includegraphics[height=40.55mm,clip=true,trim= 200 125 225 180]{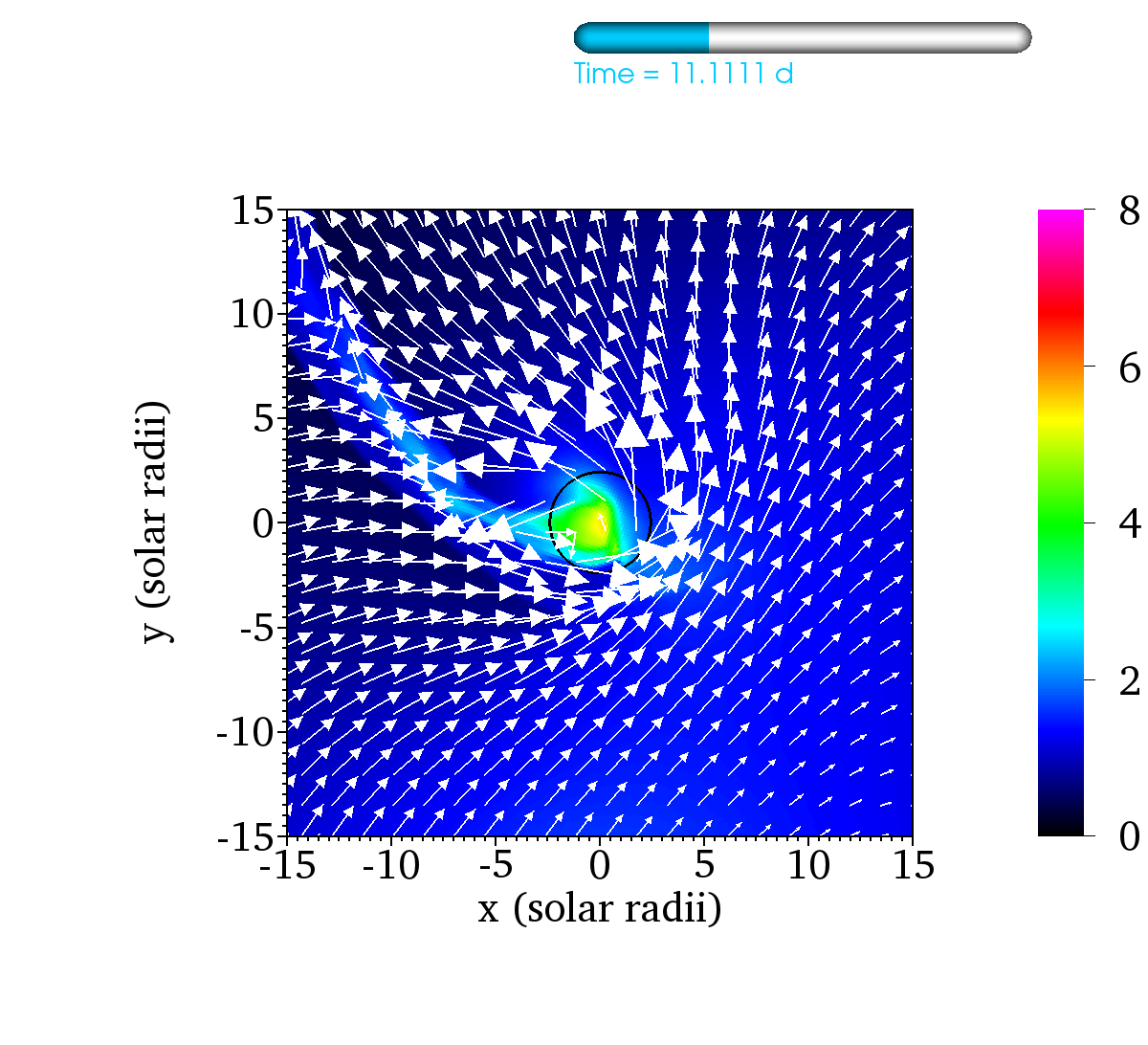}
  \includegraphics[height=40.55mm,clip=true,trim= 200 125 225 180]{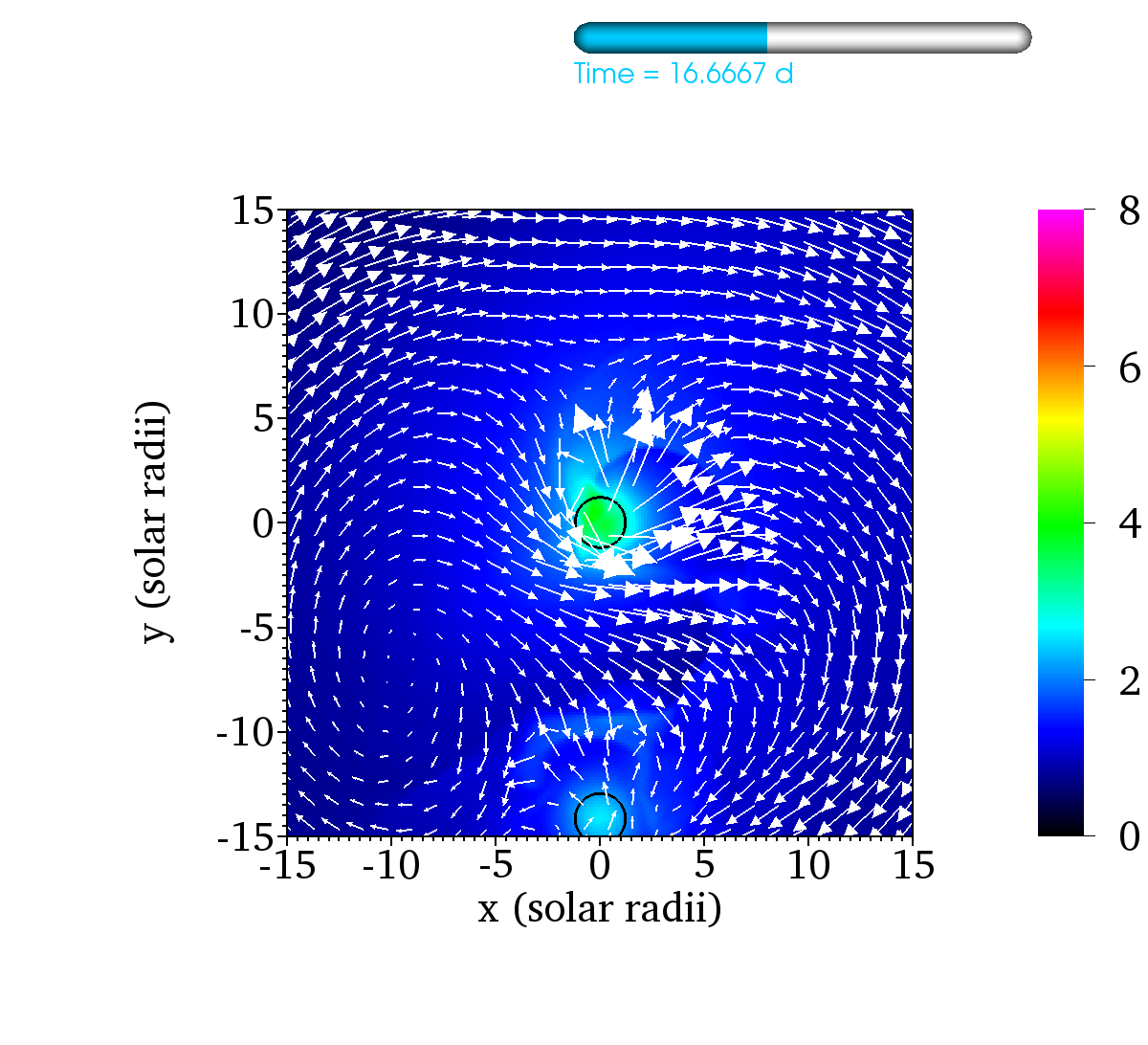}
  \includegraphics[height=40.55mm,clip=true,trim= 200 125  20 180]{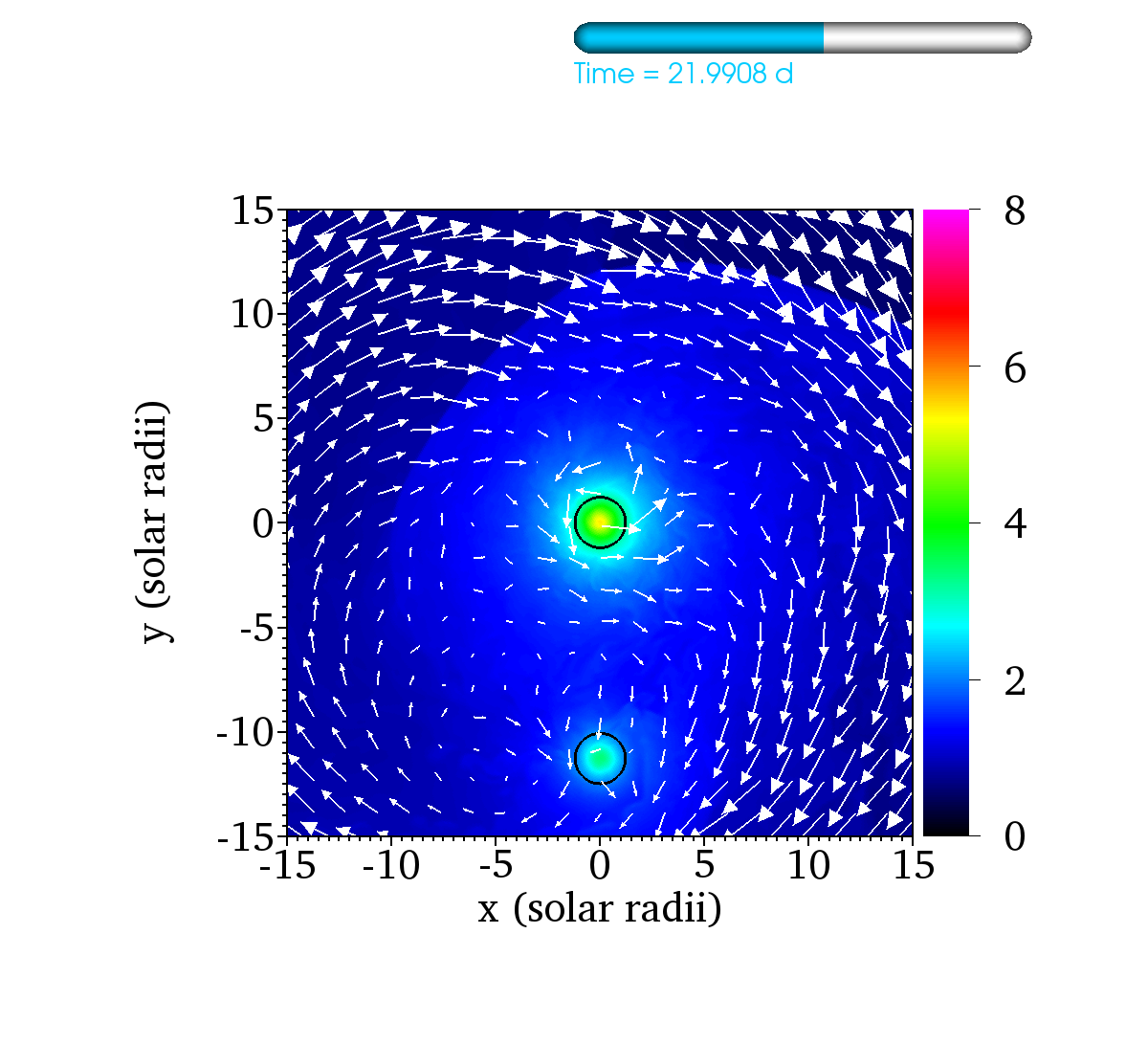}\\
  \caption{Snapshots in the orbital plane $z=0$ for Model~A.
           From left to right, columns show the times $t=6.9$, $11.1$, $16.7$ and $22.0\da$.
           Rows from top to bottom are: 
           Force density on particle~2 due to gas in the accelerating reference frame of particle~1;
           mass density normalized to $\rho\f(a)$ with velocity vectors in the corotating frame of particle~2
           (note the difference in color bar range from Fig.~\ref{fig:2D_151});
           Mach number in the corotating frame of particle~2;
           $-\phi$-component of gas velocity with respect to particle~1, in the corotating frame of particle~2,
           normalized to $v\f$, and sound speed normalized to $c\f$.
           \label{fig:2D}
          }            
\end{figure*}

Fig.~\ref{fig:2D}, shows snapshots of various quantities, sliced through $z=0$, for Model~A.
The snapshots are taken at $t=6.9$, $11.1$, $16.7$ and $22.0\da$.
Defining the end of the dynamical plunge-in phase as the time of first periastron passage \citepalias{Chamandy+19a} 
the four times represent, roughly speaking, the flow near the beginning of plunge-in, the end of plunge-in, 
 the transition to slow spiral-in, and at the beginning of slow spiral-in.
The orbital motion is counter-clockwise.
In each panel, particle~2 is at the centre and the view is rotated so that particle~1 is on the negative $y$-axis.
Circles show the softening spheres around the particles.
We focus on Model~A  because it displays the strongest deviation from theory,
and because other aspects of the run were extensively studied in Papers~I and II.

\subsection{Force density} \label{sec:force_density}
The top row shows the magnitude of the $\phi$-component 
of the force per unit volume exerted by gas on particle~2 in the co-orbiting but non-rotating rest frame of particle~1.
Positive (negative) drag contributions  are indicated by solid  (dashed) contours, spaced by the values on the color bar.
As there are positive and negative contributions from both terms in equation~\eqref{F2gas_frame1},
the plot contains four sets of contours.
Forces between gas and particle~2 dominate the contours in the upper part of the plot, 
while forces between gas and particle~1 (fictitious forces on particle~2) dominate the lower sets of contours.  A drag (thrust) on particle~1 in the lab frame produces a fictitious drag (thrust) on particle~2 in the reference frame orbiting with particle~1.
The black arrow shows the relative magnitude and direction of $\bm{F}_\mathrm{2-gas,1}$, 
while the blue arrow shows the same  for the velocity of particle~2 in the same reference frame.

At $t=6.9\da$ (column 1), 
a low density tidal tail from the primary wraps around the secondary from behind it (second row), 
providing a dynamical friction force.
The density of the gas in front of particle~2 is much smaller so the force pulling particle~2 backward dominates.
The contribution from the force on particle~1 is small because this contribution is dominated by the gas near particle~1, 
which is distributed symmetrically in the trailing and leading directions.

At $t=11.1\da$ (column 2),  the gas near particle~2 now has a higher density, leading to a greater drag force.
If logarithmic intervals in radius with respect to particle~2 contributed equally to the net force, 
as suggested by equation~\eqref{Fd},
the force per unit volume contours would  be separated by a factor $\sim10^{1/3}\approx2.15$ in radius.
 
This is roughly valid out to about the third contour:
in each of the snapshots, the ratio in radius between the innermost contours is $\sim2$, 
but the ratio between adjacent contours decreases as one moves outward from particle~2.
The ratio of radii of the fourth to third contour is generally $\sim1.5$, 
implying a decrease in contribution by a factor $\sim1.5^3/10\approx0.3$.

Calculating $R\acc$ we obtain $28$--$38\Rsun$ for $t=6.9\da$ and $15$--$16\Rsun$ for $t=11.1\da$;
the first value uses equation~\eqref{Ra} with $c_\infty$ and $v_\infty$ replaced by $c\f$ and $v\f$,
while the second value uses $|\bfv\2-\bfv\1|$ directly from the simulation.
These values correspond to roughly the third contour, so adopting $r_\mathrm{max}=R\acc$ is reasonable; the  contribution to the force beyond this radius is small.

By the next snapshot, at $t=16.7\da$ (column~3), 
the force density about particle~2 has become much more symmetric,
and the drag force is approximately zero (consistent with the top-left panel of Fig.~\ref{fig:torque_luminosity}).
Correspondingly, the contours show a high degree of left-right symmetry.
The force $\bm{F}_\mathrm{2-gas,1}$ now has comparable contributions from $\bm{F}_\mathrm{2-gas}$ 
and $-(M\2/M_\mathrm{1,c})\bm{F}_\mathrm{1-gas}$.

For the final snapshot  at $t=22.0\da$ (column 4), the $\phi$-component of $\bm{F}_\mathrm{2-gas,1}$ is quasi-steady. 
The force remains small, owing to the high degree of symmetry in the force density, 
so that the thrust and drag contributions balance except for a small net drag.
The force density pattern is quasi-steady thereafter, consistent with the net force being quasi-steady.

\subsection{Gas density evolves toward symmetry} \label{sec:normalized_gas_density}
The second row of Fig.~\ref{fig:2D} shows the gas density normalized to $\rho\f(a)$ 
of the initial density profile of the RGB star. 
Vectors show the gas velocity in the frame co-orbiting and co-rotating with particle~2,
as this is the appropriate reference frame for comparison with theory and local simulations.
The second to fifth rows are zoomed in by a factor of four compared to the top row.
At $t=6.9\da$, the density is  $\sim 10^3$ times larger than $\rho\f(a)$ 
as deeper gas is pulled tidally around particle~2.
The value of $\rho(a)/\rho\f(a)$ then reduces to $\sim10$ at $t=22.0\da$ and $\sim1$ by $t=40\da$. 
The flow around particle~2 becomes  more axisymmetric, 
rotating $\sim 20\%$ of the Keplerian value at late times \citepalias{Chamandy+18}. 
\subsection{Mach number and turbulence} \label{sec:Mach_number}
In the third row, we plot the gas Mach number computed  in the frame co-orbiting and co-rotating with particle~2.
In this frame, the gas around particle~2 is mostly supersonic during plunge-in, except in the bow shock.
For these snapshots we obtain, from earliest to latest, $\Ma\f=v\f/c\f= 5.3$, $2.1$, $1.7$, and $1.7$.
At $t=16.7\da$, a shock is still seen, now above particle~2 in the plot.
By $t=22.0\da$,  the gas about the particles and within the orbit is not only subsonic, but turbulent.
The turbulence is  well-developed by $t=18\da$.

To verify that the turbulence is not produced by the sudden change in softening length and resolution at $t=16.7\da$, 
we compared the 2D snapshots to a run (Model~F of \citetalias{Chamandy+19a}) for which the softening length and maximum AMR level are not changed in this way.
Turbulent eddies are conspicuous by $t=18\da$ in that run as well, 
even though the smallest scales of the turbulence are larger.

The onset of turbulence roughly coincides with the transition from plunge-in to slow spiral-in once $R\acc\sim a$, 
as shown in Fig.~\ref{fig:Ra}, and the particles have completed a full orbit since the first periastron passage.
The gas they encounter no longer moves out supersonically due to the deeper potential  
and confinement by overlying layers \citepalias{Chamandy+19a} so is continually being `reprocessed'.

\subsection{Azimuthal velocity and sound speed}
Finally, we present plots of the velocity and sound speed, 
which were used above to obtain modified parameter values for theoretical estimates (Sec.~\ref{sec:O99}).
The vectors in the fourth row of Fig.~\ref{fig:2D} show the gas velocity 
in the corotating frame of particle~2 as in the third row. 
The color shows the $-\phi$-component of this velocity with respect to particle~1, normalized to $v\f$.
A value of unity (orange) corresponds to the relative tangential velocity 
estimated from the initial stationary envelope profile,
while negative values (yellow to purple) denote oppositely moving gas.
In all  snapshots, the streamlines curl counter-clockwise around particle~2 
so that the $\phi$-component of the velocity reverses sign.  
The speed of gas approaching particle~2 is of order $v\f$ in the first three snapshots,
but only about $\tfrac{1}{2}v\f$ by the fourth snapshot (Sec.~\ref{sec:O99}).

Finally, in the fifth row we plot the sound speed normalized to $c\f$, 
along with the same velocity vectors plotted in rows 2 and 4.
The gas flowing toward and  deflected by the secondary has sound speed $\sim 2c\f$ (blue).

\section{Comparison to Wind Tunnel Simulations}\label{sec:comparison_windtunnel}

\subsection{Drag force comparison}
Here we compare our results with the local CE wind tunnel simulations of \citet{Macleod+17} 
(hereafter \citetalias{Macleod+17}) for which a particle representing the secondary was fixed at the centre of the grid 
and a wind was launched from the $-x$ boundary with a prescribed $x$-velocity and density gradient in the $-y$ direction.
By approximating the gas as a polytrope and assuming that the upstream wind velocity 
equals the local Keplerian orbital speed, 
the upstream Mach number is determined once the dimensionless density gradient parameter 
\begin{equation}
  \eps_\rho= \frac{2\Gn M\2}{v_\infty^2 H_\rho},
\end{equation}
and the mass ratio $q\enc=M\2/m\1(a)$ are specified.
\citetalias{Macleod+17} chose $q\enc=0.1$ and explored the dependence on $\epsilon_\rho$.
In our simulations, $q\enc=q=1/2$, $1/4$ or $1/8$ at $t=0$, but then increases with time as $m\1(a)$ decreases:
computing $m_{1,0}(a)$ from the initial envelope profile, we obtain $q_\mathrm{enc,0}= 2.0$, $1.2$ and $0.6$ at $t=40\da$ for Models~A, B and C, respectively.
\citetalias{Macleod+17} results are likely to be sensitive to their fixed choice of $q\enc$;
nevertheless we proceed with the comparison for Model~C, which proves to be fruitful.

We first  fit the drag force in \citetalias{Macleod+17} for $\gamma=5/3$ in their Fig.~10.
Replacing $\rho_\infty$ and $v_\infty$ by $\rho\f$ and $v\f$ we obtain
\begin{equation}
  \label{F_M17}
  F_\mathrm{M17}\simeq 0.6\Exp{0.61\eps_\rho}\times\frac{4\uppi\Gn^2M\2^2\rho\f}{v\f^2}.
\end{equation}
We apply this for $0.4<\eps_\rho<2.7$ (consistent with the range of parameter space explored by \citetalias{Macleod+17}) 
and plot the resulting force only for times in our simulation 
when $\eps_{\rho,0}=2\Gn M\2/(v\f^2H_{\rho,0})$ is within this range. 
The result is the dashed green line in the bottom panel of Fig.~\ref{fig:f2theory}.

At intermediate times, between  $t=12$--$26\da$, the agreement is excellent.
At $t=26\da$, when $-\dot{a}$ attains its second maximum (the first having occurred between $t=13$--$14\da$),
the $\phi$-component of $-\bm{F}_\mathrm{2-gas,1}$ decreases from its peak value,
but equation~\eqref{F_M17} predicts the force to continue rising.
At this time, $q_\mathrm{enc,0}=0.28$, or almost three times larger than that assumed by \citetalias{Macleod+17},
which likely contributes to this discrepancy.

\begin{figure}
  \begin{center}
  \includegraphics[width=70mm,clip=true,trim= 120 165 20 180]{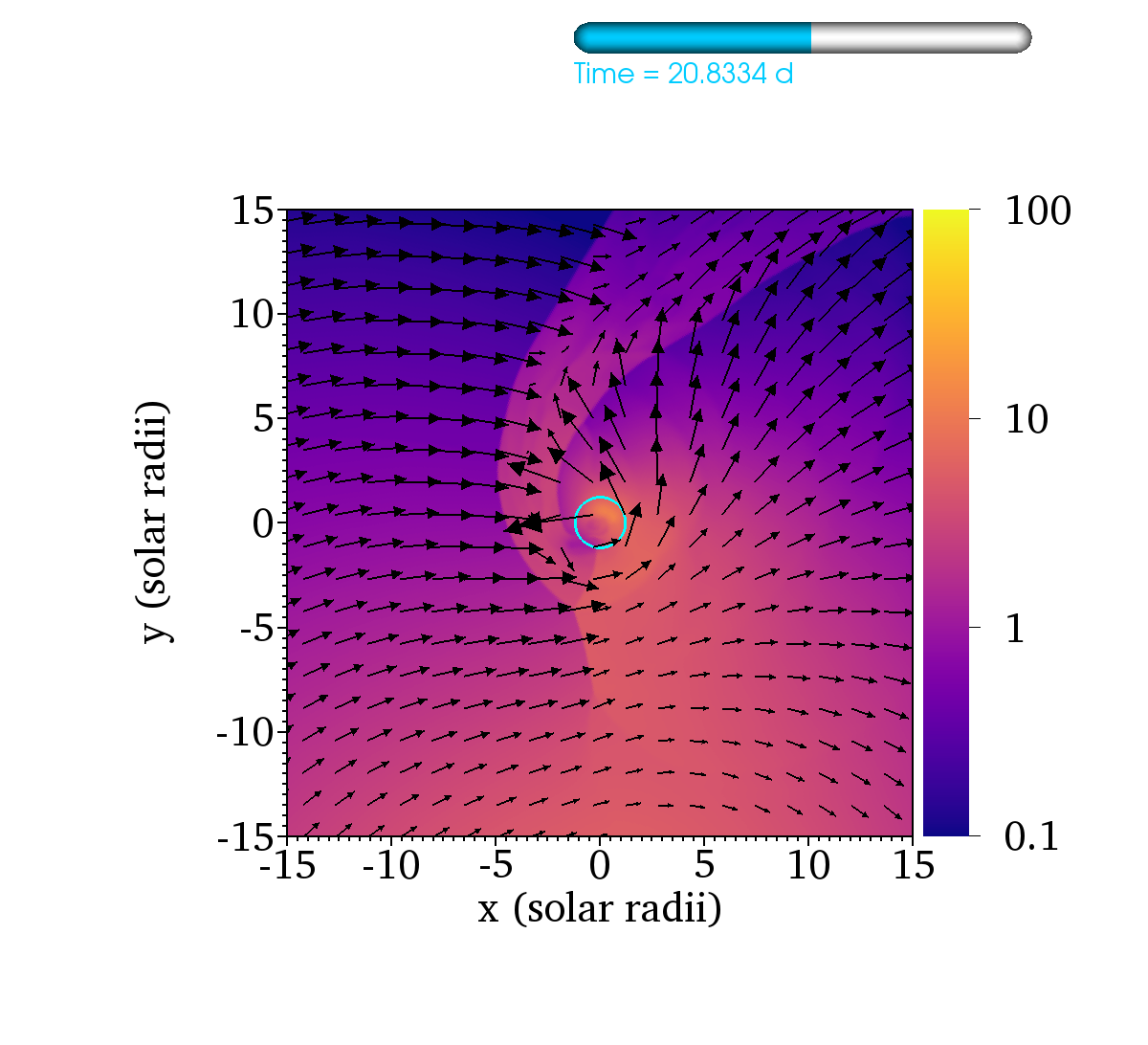}\\
  \includegraphics[width=70mm,clip=true,trim= 120 125 20 180]{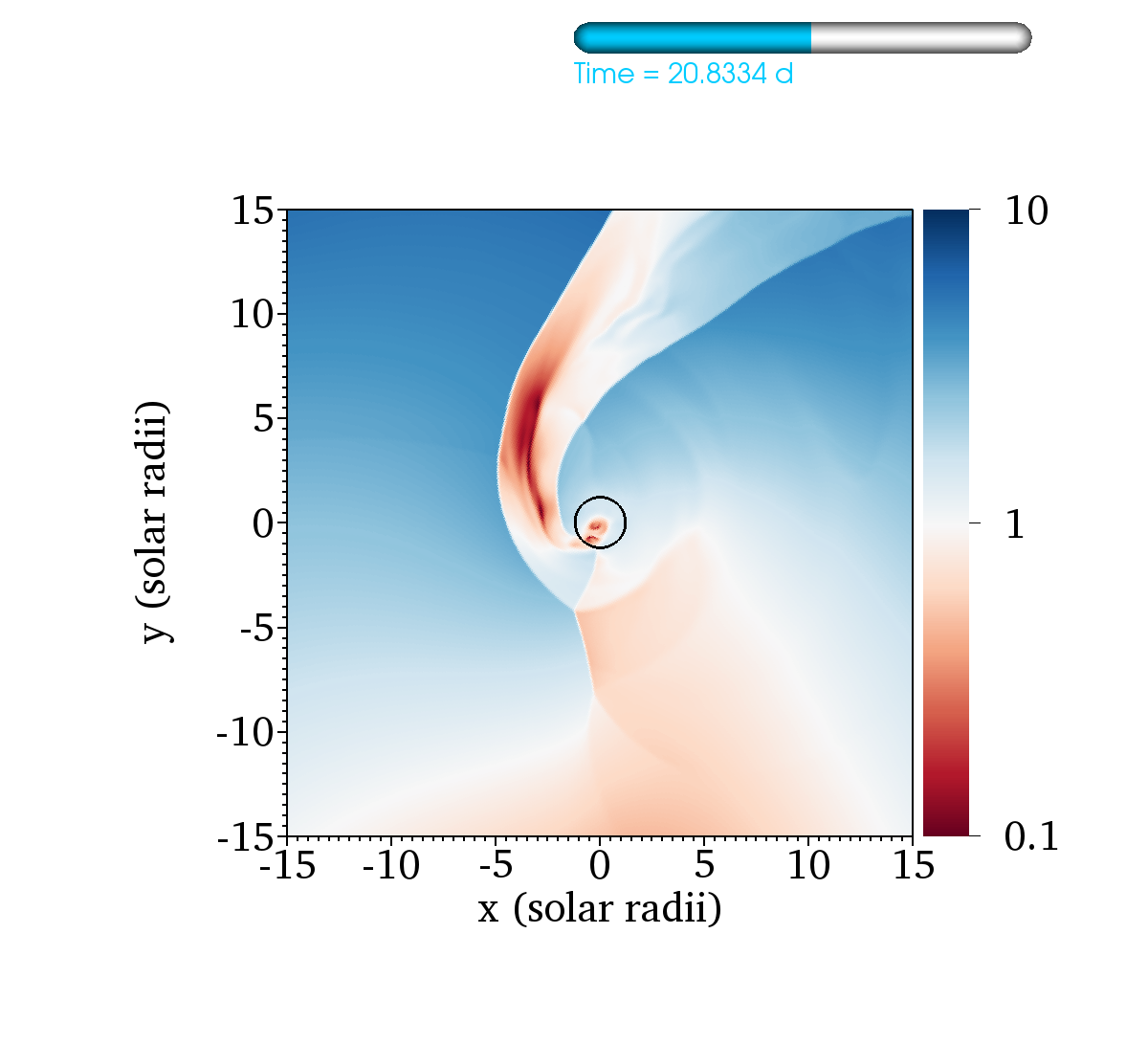}\\
  \end{center}
  \caption{Top: Slice through the orbital plane of density normalized to $\rho\f(a)$ along with velocity vectors 
           in the reference frame orbiting and corotating with particle~2 for Model~C at $t=20.8\da$,
           when $\eps_\rho=0.80$ and $q\enc=0.15$. 
           Bottom: Mach number in the same reference frame.
           At this time, the drag force is approximately equal to that predicted from the fitting formula of \citetalias{Macleod+17}.
           These snapshots can be compared with those from Fig.~2 of \citetalias{Macleod+17},
           keeping in mind that the unit of their axes is $2\Gn M\2/v\f^2=7.4\Rsun$.
           There is a close correspondence between the global and local simulations, as expected.
           \label{fig:2D_151}
           }
\end{figure}

\subsection{Flow structure comparison}
Flow structure of global and local simulations can also be compared.
For Model~C, we choose the time $t=20.8\da$, at which $\eps_\rho=0.80$ (and $q\enc=0.15$), 
to compare with the lower left panels of Fig.~2 of \citetalias{Macleod+17}.
We plot the mass density normalized to $\rho\f(a)$ and velocity vectors in the corotating frame of particle~2
in the top panel of Fig.~\ref{fig:2D_151}, and the Mach number in the corotating frame of particle~2 in the bottom panel.
The unit of the \citetalias{Macleod+17} axes is $2\Gn M\2/v\f^2=7.4\Rsun$, so our plotting region is slightly larger than theirs.
The level of agreement is  remarkable. 
All of this suggests that  the local simulations approximate global simulations for this window of parameter space.

Curiously, there is also some correspondence between the flow in Model~A and that of the local simulations,
even though the mass ratio of the former is much larger.
Whereas \citetalias{Macleod+17} adopted $q\enc=0.1$ and $\eps_\rho=0.2$--$2$,
Model~A has $q_\mathrm{enc,0}=0.50$, $0.75$, $1.23$ and $1.53$ 
and $\eps_{\rho,0}=20.3$, $2.1$, $1.7$ and $1.4$ at
$t=6.9$, $11.1$, $16.7$ and $22.0\da$, respectively.
Despite differences in $q\enc$ and the force seen in Fig.~\ref{fig:f2theory},
the panels of Fig.~\ref{fig:2D} showing $\rho/\rho\f$ and $\Ma$ in Model~A at $11.1\da$
show  similarities 
to those of $\eps_\rho=2.00$ in Fig.~2 of \citetalias{Macleod+17},
as seen by comparing the second column, third and fourth rows of Fig.~\ref{fig:2D} 
with Fig.~2 of \citetalias{Macleod+17}.
The flow pattern is  similar and both methods exhibit a thin spiral shock.  
However,  the normalized gas density in Fig.~\ref{fig:2D} is $1$--$1000$
(whereas in Fig.~\ref{fig:2D_151} we used $0.1$--$100$, as in \citetalias{Macleod+17}).
Thus {in Model~A} our normalized densities are almost an order of magnitude larger than those of \citetalias{Macleod+17}, 
likely because our $q\enc$ is $7.5$ times larger than theirs at that time.
The size of the region plotted in units of $2\Gn M\2/v_\infty^2$ differs from \citetalias{Macleod+17}: 
for Model~A we obtain $2\Gn M\2/v\f^2=29$, $19$, $16$ and $14\Rsun$, respectively, for the four snapshots.\footnote{
When $\eps_\rho$ and $2\Gn M\2/v_\infty^2$ are estimated using the actual velocity $\bfv\2-\bfv\1$, rather than $v\f$, 
the values are larger by $37\%$ for $t=6.9\da$ but hardly differ for the other snapshots of Model~A.}

As expected, snapshots  at other times hardly resemble those of \citetalias{Macleod+17}.
At $t=6.9\da$, $\eps_\rho$ is an order of magnitude larger than that explored by \citetalias{Macleod+17}.
We do see  increasing density contrast and larger rotation angle of the bow shock with increasing $\eps_\rho$,
as in \citetalias{Macleod+17}, but our shock  is thick and  morphologically complex.
By $t=16.7\da$, $R\acc$ has already become comparable to $a$, as shown in the top panel of Fig.~\ref{fig:Ra}.
The assumptions of \citetalias{Macleod+17}, namely
that (i) the envelope gas encountered by the secondary had not been previously affected; 
(ii) their $\rho_\infty$ smoothly and monotonically decreases with distance from the RGB core, 
and (iii) the gravity force from the RGB core can be approximated as everywhere downward,
are no longer valid.
Moreover, in our final snapshot, turbulence likely affects the dynamics.

Thus, we would not expect wind tunnel simulations 
to approximate  the results of Model~A at \textit{late times} even if several wind tunnel simulations of different $q_\mathrm{enc,0}$ and $\eps_{\rho,0}$ were patched together to accommodate dynamically changing values of these parameters in the global simulation.
However, given the excellent agreement at intermediate times for Model~C, 
it would be interesting to 
compare local and global simulations using such dynamical patching of the local simulations
to refine the temporal range over which this approach
could  be useful and computationally efficient.

\section{How Important is Radiative Transfer?}
Radiative transfer is neglected in our simulations (and in virtually all global CE simulations to date).
A large diffusive flux of radiation out of the central region might lead to a different flow structure at late times.
Averting the buildup of thermal energy might allow the flow there to retain a structure closer to that which it had originally,
and thus closer to that assumed in wind tunnel experiments,
where the flow is assumed to be unaffected by previous orbital passages of the particles.

Thus, we estimate the diffusion time to determine whether cooling would be significant at late times in our simulations.
We consider the flow properties at $t=22.0\da$ in Model~A (right column of Fig.~\ref{fig:2D}), 
focussing on the region around particle~2 with $\Ma<1$.
The distance from particle~2 to the boundary of this region is estimated as $R\sim7\Rsun$.
Within this region, a typical gas density is $\rho\sim2\times10^{-4}\gcmcmcm$ and the temperature $T>10^6\K$.
At this temperature, hydrogen gas is ionized and the opacity is dominated by electron scattering,
with cross section $\sigma_\mathrm{T}=6.65\times10^{-25}\cm^2$.

The diffusion time is estimated by multiplying the number of scatterings by the mean free path and dividing by the speed of light in vacuum $c$.
The mean free path is given by $l= (n_\mathrm{e}\sigma_\mathrm{T})^{-1}$, 
the electron number density by $n_\mathrm{e}=\rho/m_\mathrm{H}$ where $m_\mathrm{H}$ is the mass of the hydrogen atom,
and the number of scatterings is given by $N\simeq R^2/l^2$.
Thus, we have
\begin{equation}
  t_\mathrm{d}\sim \frac{Nl}{c} \sim \frac{R^2\rho\sigma_\mathrm{T}}{m_\mathrm{H}c},
\end{equation}
or $t_\mathrm{d}\sim20\yr$ for Model~A (with corresponding optical depth $\tau\sim\sigma_\mathrm{T}n_\mathrm{e}R\sim40$).
For Models~B and C, at a comparable time in the evolution, $R$ is somewhat smaller while $\rho$ is slightly larger than in Model~A,
and the value of $t_\mathrm{d}$ is of the same order of magnitude.
Since $t_\mathrm{d}\gg t$, the neglect of radiative transfer in the $\Ma<1$ region around the particles is justified.

\section{Conclusions} \label{sec:conclusions}
We computed the drag force in three global CE simulation runs of $40\da$ 
in which a companion point particle is placed in circular orbit around a 2$\Msun$ RGB star. 
The runs are identical except for the value of the companion mass, 
$\Msun$, $\tfrac{1}{2}\Msun$ or $\tfrac{1}{4}\Msun$.
We found that:
\begin{itemize}
  \item The drag force on the particles at late times, during the slow spiral-in phase, 
        has mean magnitude $\sim7\times10^{33}\dyn$, depending only weakly on companion mass, 
        and varies periodically with the orbit (Figs.~\ref{fig:force2}, \ref{fig:force2_comparison}).
  \item BHL/DM theory overestimates the drag force at late times
        by at least an order of magnitude for the run with initial mass ratio $q=1/2$ (Fig.~\ref{fig:f2theory} top panel), 
        and cannot reproduce the late time force for any of the three runs.
  \item BHL/DM theory and local wind tunnel simulations are particularly inapplicable at late times for large $q\enc=M\2/m\1(a)$ 
        because the accretion radius becomes comparable to the inter-particle separation. 
        The gas encountered by the particles forms a turbulent, thermalized, 
        highly symmetric region around the particles (Fig.~\ref{fig:2D} rightmost column).
        Hydrodynamic drag may even dominate over dynamical friction during this phase, but further work is needed. 
  \item At earlier times, the drag force peaks at or just before the first periastron passage 
        with value approximately proportional to the companion mass (Fig.~\ref{fig:force2}). 
        Near this peak, the drag force  is reasonably  well matched by BHL/DM theory 
        and particularly well matched by local wind tunnel simulations (Fig.~\ref{fig:f2theory} bottom panel), 
        which also reproduce various features of the 2D slices at that time 
        (c.f. Fig.~\ref{fig:2D_151} of this work and Fig.~2 of \citetalias{Macleod+17}).
\end{itemize}

Thus, for low $q\enc$, BHL/DM theory and local wind tunnel simulations 
approximate the drag in global simulations during the intermediate plunge-in phase, but not before or after.
Since $q\enc$ evolves temporally in global simulations,  
different fixed $q\enc$ wind tunnel simulations must be patched together to increase the fidelity 
of comparison with global simulations over a larger temporal range. 
This has not yet been done.

Finally, more general theoretical approaches are needed to account for the 
high degree of symmetry and turbulence 
in the flow once $R\acc \sim a$, and the associated reduced drag at late times.
This reduced drag dramatically slows the inward evolution 
and explains why numerous CE simulations do not reach tight enough orbits by the end of runs to eject the CE envelope.

\section*{Acknowledgements}
We thank Orsola De~Marco, Rosa Everson, Robert Fisher, Morgan MacLeod, Jason Nordhaus, Enrico Ramirez-Ruiz, and Jan Staff for useful discussions. 
We are grateful to the referee for a helpful report.
EB acknowledges the Aspen Center for Physics, supported by NSF grant PHY-160761.
This work used the computational and visualization resources in the Center for Integrated Research Computing (CIRC) at the University of Rochester and the computational resources of the Texas Advanced Computing Center (TACC) at The University of Texas at Austin, provided through allocation TG-AST120060 from the Extreme Science and Engineering Discovery Environment (XSEDE) \citep{xsede}, which is supported by National Science Foundation grant number ACI-1548562. Financial support for this project was provided by the Department of Energy grant DE-SC0001063, the National Science Foundation grants AST-1515648 and AST-181329, and the Space Telescope Science Institute grant HST-AR-12832.01-A.
\appendix

\section{Effect of changing softening length and resolution} \label{sec:152}
Here we compare the $-\phi$-component of the force exerted on particle~2 by the gas in the frame of particle~1
in Model~A and Model~F of \citetalias{Chamandy+19a}.
Model~F restarts from Model~A at $t=16.7\da$ but the softening radius and smallest resolution element are not halved as in Model~A.
The evolution of the force is very similar, confirming that the halving of $r\soft$ and $\delta$ does not importantly affect the overall evolution of the force.

\footnotesize{
\noindent
\bibliographystyle{mnras}
\bibliography{refs}
}

\begin{figure*}
\begin{center}
  \includegraphics[width=0.8\textwidth,clip=true,trim= 0 0 0 0]{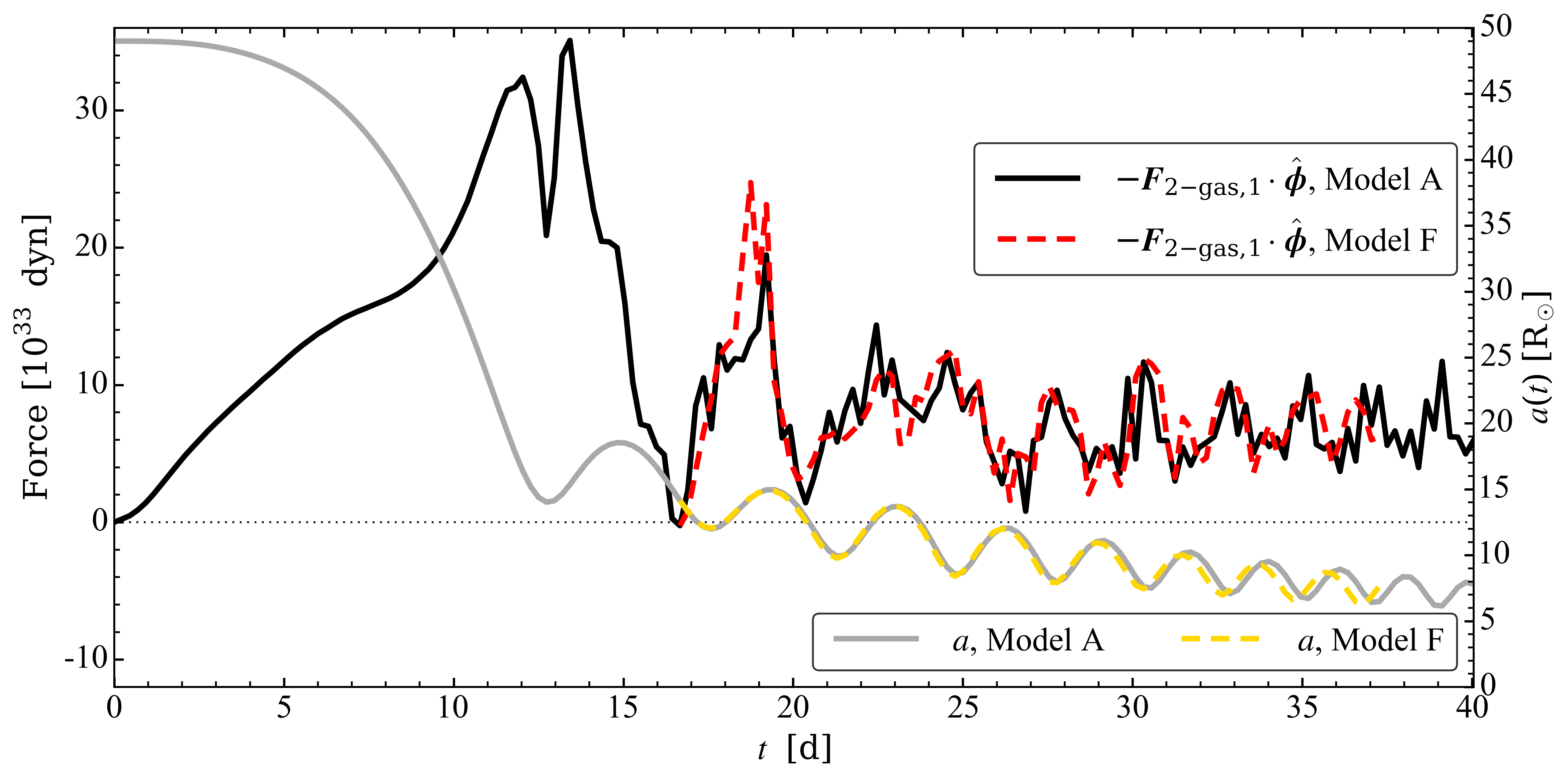}
  \caption{Comparion between the $\phi$-component of the force exerted by gas on particle~2 in the reference frame of particle~1,
           for Model~A and Model~F of \citetalias{Chamandy+19a}. 
           In Model~F, the softening radius and smallest resolution element 
           were kept constant during the simulation rather than being halved at $t=16.7\da$, as in Models~A, B and C. 
           \label{fig:force2_comparison}
          }            
\end{center}
\end{figure*}

\end{document}